\documentclass[9pt,twocolumn,twoside]{pnas-new} 

\usepackage{hyperref}
\usepackage{cleveref}

\templatetype{pnasresearcharticle} 
\setboolean{displaywatermark}{false}

\title{Bayesian regression explains how human participants handle parameter uncertainty}
\author[a,b,2]{Jannes Jegminat}
\author[c,d]{Maya A. Jastrzębowska}
\author[c,e]{Matthew V. Pachai}
\author[c]{Michael H. Herzog}
\author[a,b]{Jean-Pascal Pfister}
\affil[a]{Department of Physiology, University of Bern, 3012 Bern, Switzerland}
\affil[b]{Institute of Neuroinformatics and Neuroscience Center Zurich, ETH and the University of Zurich, 8057 Zurich, Switzerland}
\affil[c]{Laboratory of Psychophysics (LPSY), Brain Mind Institute, School of Life Sciences, École Polytechnique Fédérale de Lausanne (EPFL), 1015 Lausanne, Switzerland}
\affil[d]{Laboratory for Research in Neuroimaging (LREN), Department of Clinical Neuroscience, Lausanne University Hospital and University of Lausanne, 1011 Lausanne, Switzerland}
\affil[e]{Department of Psychology, York University, ON M3J 1P3 North York, Canada}

\leadauthor{Jegminat} 

\significancestatement{A central function of the brain is learning a model of the world from sensory inputs. Taking into account the noise and ambiguity of these inputs, the brain works with levels of belief rather than single values. Previous research mostly focused on how the brain represents its beliefs about the outside world. Here, we ask the following question: how do different levels of belief influence predictions made by the brain's model? We consider three methods of doing this, ranging from "not at all" to "totally", and compare them to human performance in a visual experiment. Our results suggest that brains are well aware of their degree of belief when making predictions.}

\authorcontributions{J.J., M.J., M.P. M.H. and J.P. designed research; M.J., M.P., M.H. conducted the experiment; J.J., J.P. analyzed data; and J.J., M.J., M.H., J.P. wrote the paper.
}
\authordeclaration{The authors declare no conflict of interest here.}
\correspondingauthor{\textsuperscript{2}To whom correspondence should be addressed. E-mail: jannes@ini\@uzh.ch}


\begin{abstract}
The human brain copes with sensory uncertainty in accordance with Bayes' rule. However, it is unknown how the brain makes predictions in the presence of parameter uncertainty. Here, we tested whether and how humans take parameter uncertainty into account in a regression task. Participants extrapolated a parabola from a limited number of noisy points, shown on a computer screen. The quadratic parameter was drawn from a prior distribution, unknown to the observers. We tested whether human observers take full advantage of the given information, including the likelihood function of the observed points and the prior distribution of the quadratic parameter. We compared human performance with Bayesian regression, which is the (Bayes) optimal solution to this problem, and three sub-optimal models, namely maximum likelihood regression, prior regression and maximum a posteriori regression, which are simpler to compute. Our results clearly show that humans use Bayesian regression.
We further investigated several variants of Bayesian regression models depending on how the generative noise is treated and found that participants act in line with the more sophisticated version.

\end{abstract}

\dates{This manuscript was compiled on \today}
\doi{\url{www.pnas.org/cgi/doi/10.1073/pnas.XXXXXXXXXX}}

\begin{document}

\maketitle
\thispagestyle{firststyle}
\ifthenelse{\boolean{shortarticle}}{\ifthenelse{\boolean{singlecolumn}}{\abscontentformatted}{\abscontent}}{}

\dropcap{T}he brain evolved in an environment that requires fast decisions to be made based on noisy, ambiguous and sparse sensory information, noisy information processing and noisy effectors. Hence, decisions are typically made under substantial uncertainty. The Bayesian brain hypothesis states that the brain uses the framework of Bayesian probabilistic computation to make optimal decisions in the presence of uncertainty \citep{knill2004bayesian,vilares2011bayesian,friston2012bayesian}. A large body of research has established that many aspects of cognition are indeed well described by Bayesian statistics. These include magnitude estimation \citep{petzschner2011magnitudeestimation}, color discrimination \citep{olkkonen2014}, cue combination \citep{ernst2002humans}, cross-modal integration \citep{Alais2004,Shams2010}, integration of prior knowledge \citep{berniker2010learning,girshick2011cardinal} and motor control \citep{kording2004bayesian,trommershauser2005optimal,landy2012dynamic}.

Most of these experimental studies can be cast into the problem of estimating a hidden quantity from sensory input. Much fewer experimental studies have been performed on regression tasks (but see \citep{lucas2015rational} for an overview, and, e.g., \citep{carroll1963functional,delosh1997extrapolation,villagradata2018}). In a regression task, the aim is to learn the mapping from a stimulus $x$ to an output $y$ after having been exposed to a training data set $D = \{(x_i,y_i)\}_{i=1}^N$ of $N$ associations between stimulus $x_i$ and its corresponding $y_i$. Since the mapping from $x$ to $y$ can be probabilistic, the aim of regression is to find an expression for $p(y|x,D)$. Classification tasks (such as object recognition) or self-supervised tasks such as estimating the future position of an object from past observations are just a few examples from a long list of regression tasks performed on a daily basis.

The machine learning literature contains many solutions to the regression problem, such as nonlinear regression, support vector machines, Gaussian processes or deep neural networks (see \citep{barber2012bayesian} for an introduction). It is unclear, however, how humans perform regression tasks. Most of the machine learning solutions rely on the assumption that the mapping from $x$ to $y$ is parametrized by a set of parameters $w$, such that the original regression problem of finding $p(y|x,D)$ is replaced by a parameter estimation problem, i.e., finding the best set of parameters $w^*$ for the parametrized mapping $p(y|x,w^*)$. However, this approach is not Bayesian since no uncertainty over the parameters $w$ is included in the regression model.

The Bayesian approach to regression proceeds in two steps \citep{mackay1992bayesian}. First, the posterior distribution over the parameters $p(w|D)$ is computed from the observed data $D$. Then, this posterior is used to compute the posterior predictive distribution by integrating over the parameters:
\begin{align}
p(y|x,D) = \int p(y|x,w) p(w|D) \text{d}w
\end{align}
Taking into account the uncertainty over parameters is particularly relevant for predictions when the data set size $N$ is small compared to the number of parameters. Indeed, taking into account the uncertainty helps to generalize to unknown data and thereby alleviates overfitting. Parameter uncertainty also plays a key role in computing the prediction uncertainty. Both of these aspects -- overfitting on small data sets and lack of prediction uncertainty -- currently limit the power of deep neural network models \citep{gal2016dropout,blundell2015weight}. These models have millions of parameters and their performance grows with the number of layers \citep{zagoruyko2017diracnets,huang2017densely}. To prevent overfitting, training these models requires ever larger and more expensive training sets.\\ \indent
It is interesting to note that classic DNNs do not use weight uncertainty and are therefore limited in their ability to compute the prediction uncertainty. Recently, the idea of computing the probability distribution over weights in DNNs and using it for prediction has gained traction and given rise to the so-called Bayesian Neuronal Network (BNN), for example  \citep{fortunato2017bayesian,li2017dropout}. BNNs promise better performance in the low-data regime. Additionally, BNNs output their prediction uncertainty, which is crucial when the cost of decisions is unequally distributed as is usually the case in behavioural tasks \citep{maloney2007questions,cohen2007should,jarbo2018sensory}. 
\\ \indent
Here we ask if humans take advantage of Bayesian regression. To address this question, we conducted the following psychophysical experiment. Participants sat in front of a screen on which 4 points from a noisy parabola were shown. Their task was to correctly extrapolate the parabola, i.e., to find the vertical position of the parabola at a given horizontal location. The quadratic parameters of the parabolas were drawn from a bimodal prior distribution, designed to make the parabolas face either upwards or downwards. After recording the participant's response, we showed the parabola that generated the dots as feedback. This feedback enabled the participants to learn both the prior and the generative model. Because we wanted to test to what extent participants made decisions in accordance with a Bayesian regression strategy, we varied the level of noise of the parabola. The rationale is that the higher the noise level, the higher the uncertainty about the correct parameter and, according to Bayesian regression, the more participants should rely on the prior. We found that Bayesian regression indeed explained participants' responses better than maximum likelihood regression and maximum a posteriori regression. The latter two models are widely used by applied statisticians \citep{neter1996applied,guindon2003simple,redner1984mixture,felsenstein1981evolutionary, myung2003tutorial,myers1990classical,sorenson1980parameter,gauvain1994maximum,forster2015imu} and also in psychophysical modelling \citep{acuna2015using} but fail to account for parameter uncertainty.

\begin{figure}[H] 
\begin{minipage}{0.49\textwidth}
\includegraphics[width=\textwidth]{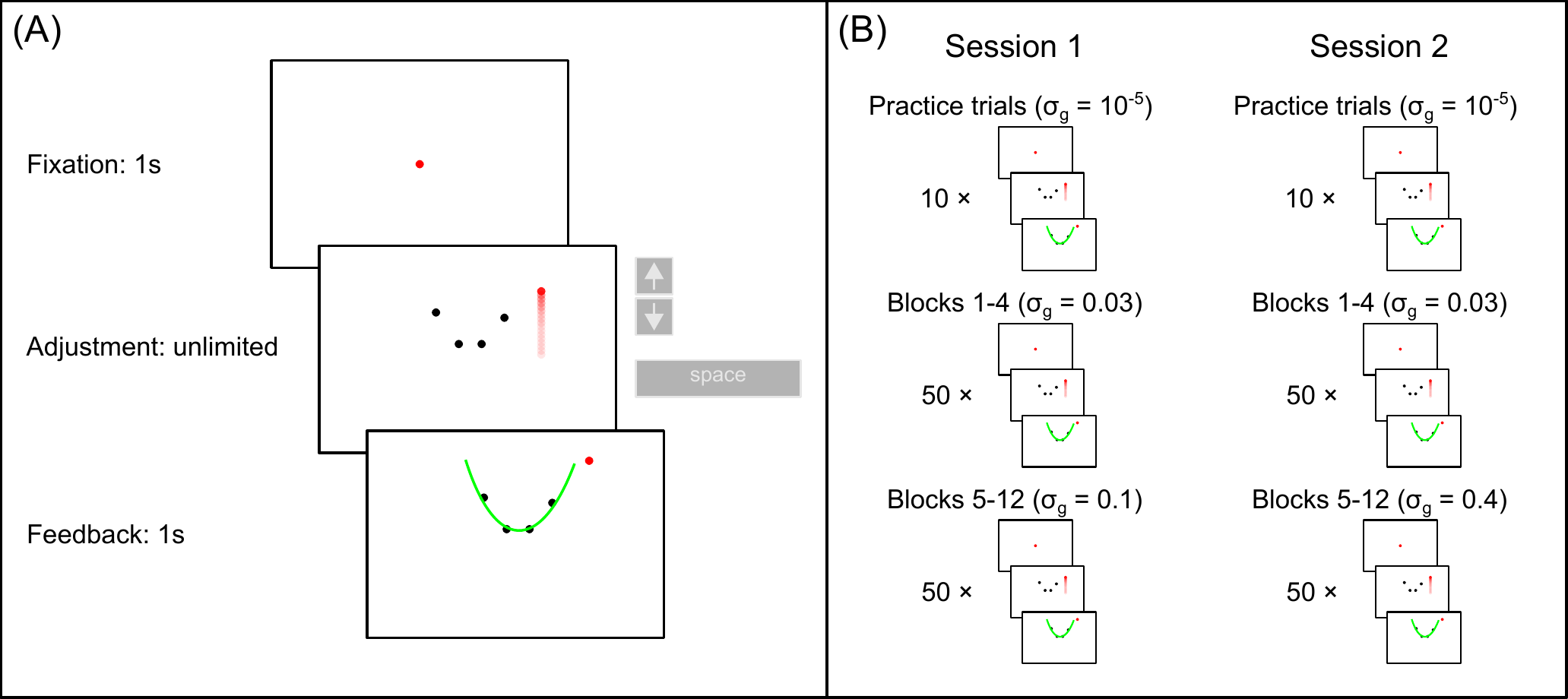}
\caption{ \label{fig:psychophysics-setup}
Experimental protocol. \textbf{(A)}: procedure of a single trial. First, a fixation dot was presented for 1s before the 4-dot stimulus appeared. Observers then had unlimited time to adjust the fifth dot with the up and down arrow keys. They then clicked the space bar to confirm the final position of the adjustable dot. After the response, the generative parabola was shown for 1s as feedback. \textbf{(B)}: experimental protocol. Both sessions began with 10 practice trials with virtually no noise ($\sigma_g = 10^{-5}$), followed by 4 blocks of 50 trials of low noise ($\sigma_g = 0.03$).  In session 1, the low noise blocks were followed by 8 blocks of 50 trials of medium noise ($\sigma_g = 0.1$), while in session 2, the low noise blocks were followed by 8 blocks of 50 trials of high noise ($\sigma_g = 0.4$). In total, each subject ran 400 trials per noise level. See Materials and Methods \Cref{sec:methods:participants} for more details.
}
\end{minipage}
\end{figure}

\section{Results}
\subsection{A novel paradigm to test regression}
In each trial, we chose the parameter $w$ of a parabola $y = wx^{2}$ from a fixed prior distribution $\pi(w)$. Participants were shown 4 dots which were on this parabola, but the parabola itself was not shown. We had 20 such 4-dot stimuli
, which we repeated 20 times each. For a certain 4-dot stimulus $D_j$, the parameter $w_j$ was always the same. The x-positions of the 4 stimulus dots were fixed across all trials and were rather close to the vertex of the parabola. The parameter $w_j$ was either positive (parabola opening upwards) or negative (parabola opening downwards), with the same probability, i.e., 0.5. The prior distribution $\pi(w)$ was bimodal with means of $\pm{1}$ and standard deviations of 0.1 (see Material and Methods \cref{sec:methods:stimulus-generation}). We did not tell observers about the prior probability distribution. Also unbeknown to the observers, we jittered the dots for each stimulus $D_j$ along the y-axis by adding zero-mean Gaussian noise of a certain level $\sigma_g$. The jitter was the same for all 20 repetitions of a stimulus $D_j$. A fifth dot was presented to the right of the 4-dot stimulus, always at the same x-position $x^\star$. Participants could move the fifth dot up and down along the y-axis by using the up and down arrow keys. Participants were asked to adjust the y-position so that the dot correctly extrapolated the parabola. After the adjustment, we showed the generating parabola and the adjusted point as feedback.

Because we wanted to test to what extent observers rely on prior information, we used low, middle and high values of $\sigma_g$. The rationale is that the higher the noise level, the higher the uncertainty (the lower the likelihood) and the more participants rely on the prior. Thus, for each participant and each of the 3 noise levels, we ran 20 4-dot stimuli with 20 repetitions, which were randomly ordered (\Cref{fig:psychophysics-setup}, right). The set of responses to a 4-dot stimulus $D_j$ is $R_j = \{r^{(j)}_1, \dots r^{(j)}_{20} \}$. The different regression models make different predictions as to the extent to which participants use the generative probability $p(y|x,w)$ and the prior $\pi(w)$.
\subsection{The regression models}
\begin{figure}[H]
\begin{minipage}{0.5\textwidth}
\begin{tabular}{ll}
{\bf (A)} & {\bf (B)} \\
\begin{minipage}{0.45\textwidth}
\includegraphics[width=\textwidth,trim={0cm 0cm 0cm 0cm},clip]{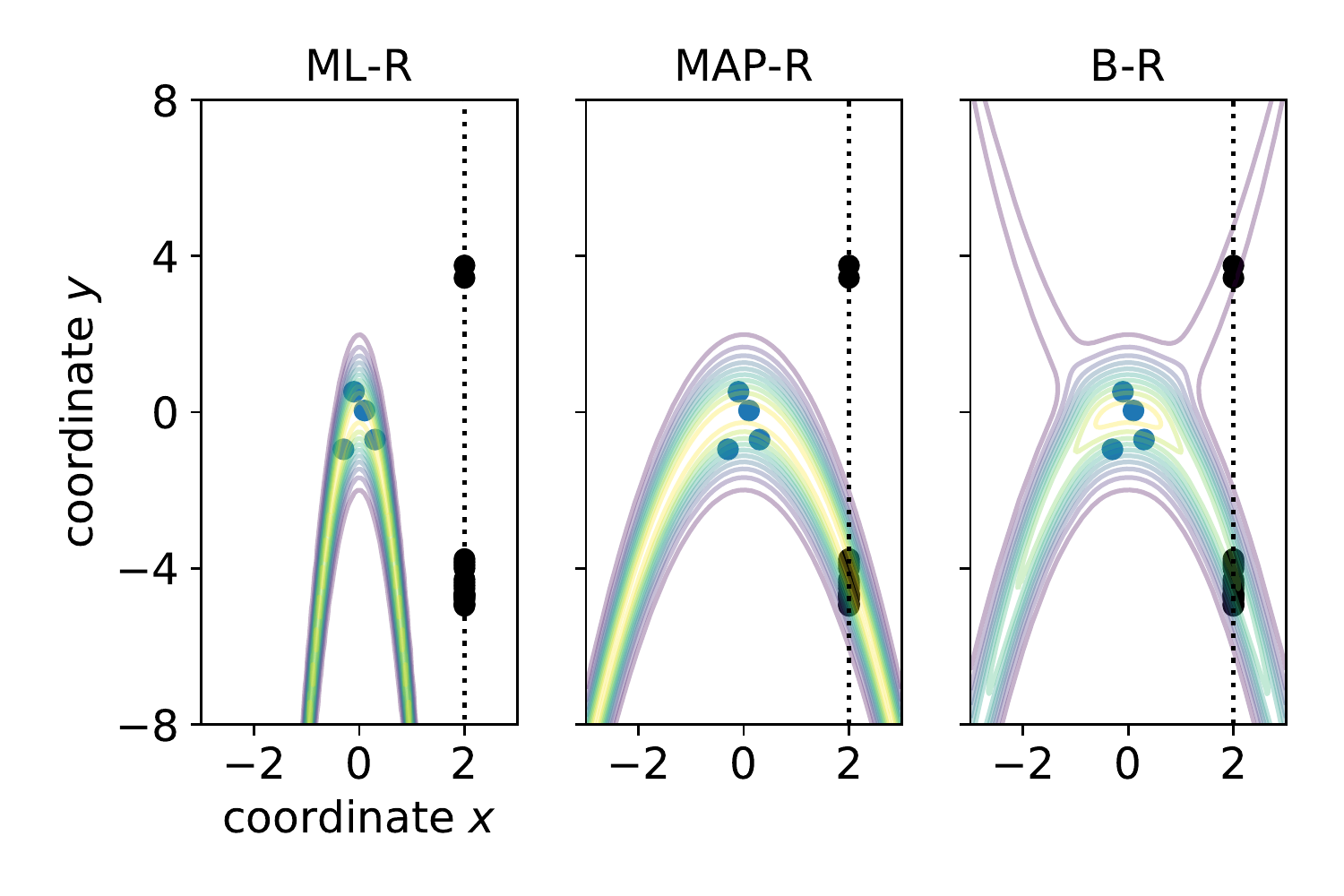}
\end{minipage} &
\begin{minipage}{0.45\textwidth}
\includegraphics[width=\textwidth]{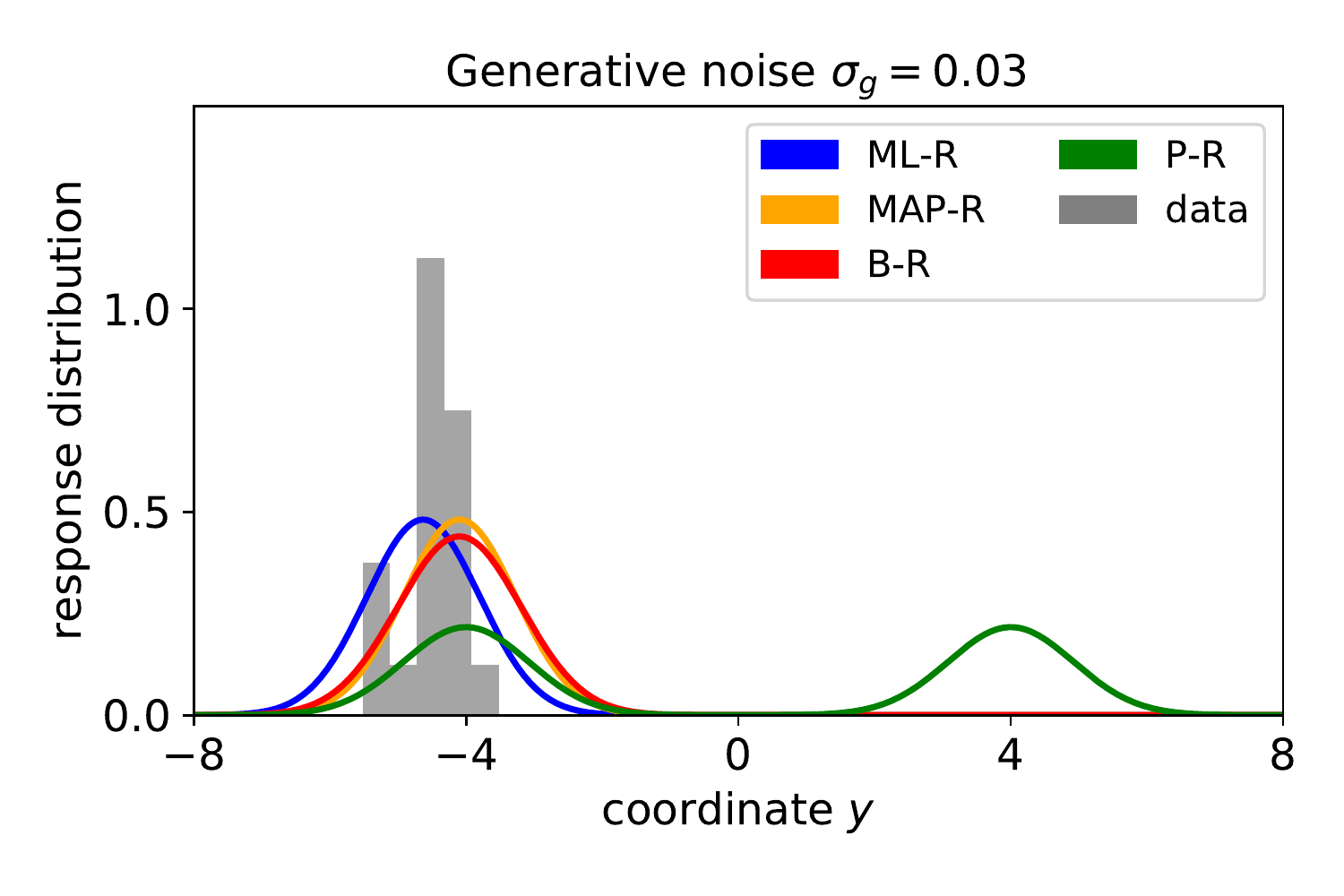}
\end{minipage} \\
{\bf (C)} & {\bf (D)} \\
\begin{minipage}{0.45\textwidth}
\includegraphics[width=\textwidth]{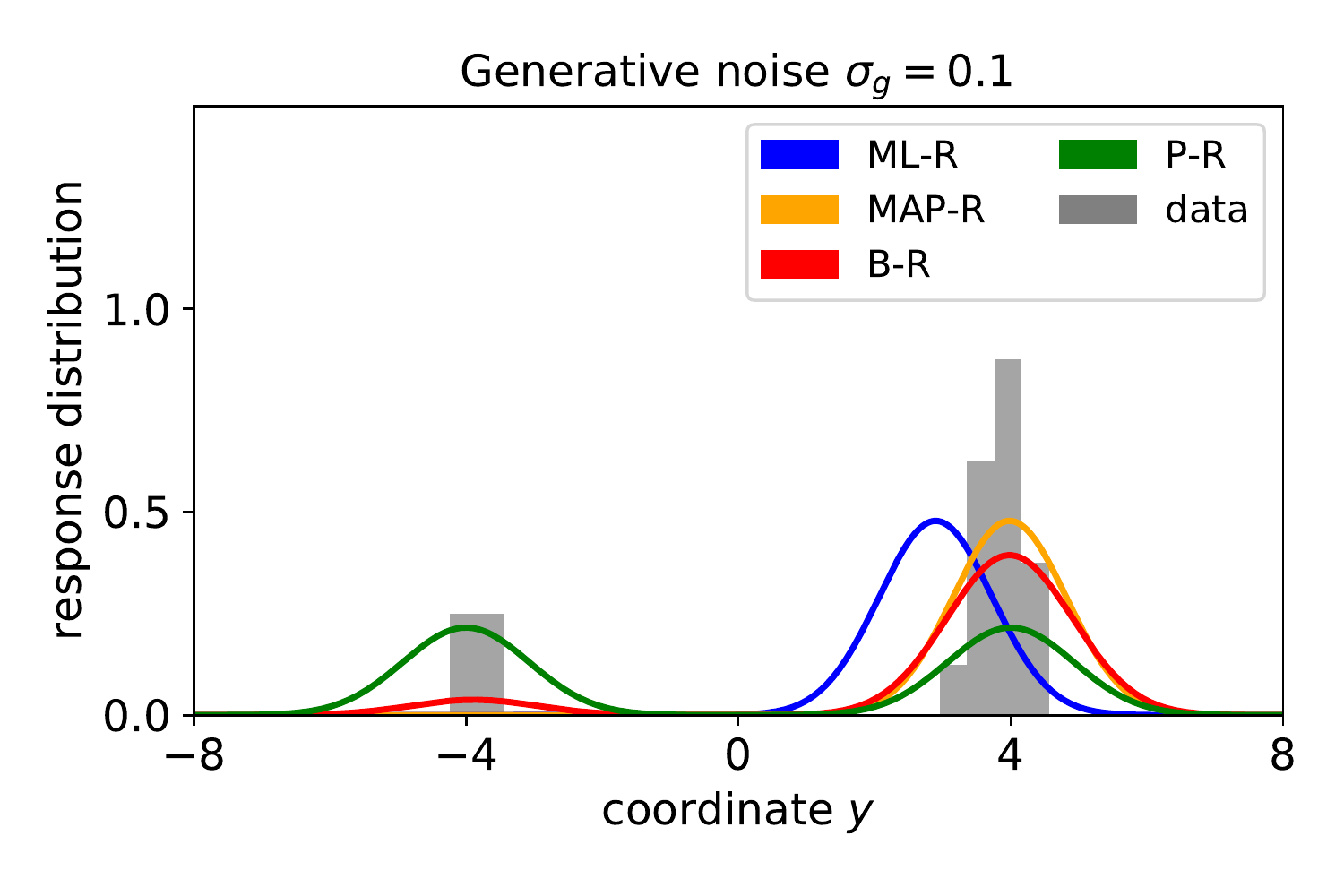}
\end{minipage} &
\begin{minipage}{0.45\textwidth}
\includegraphics[width=\textwidth]{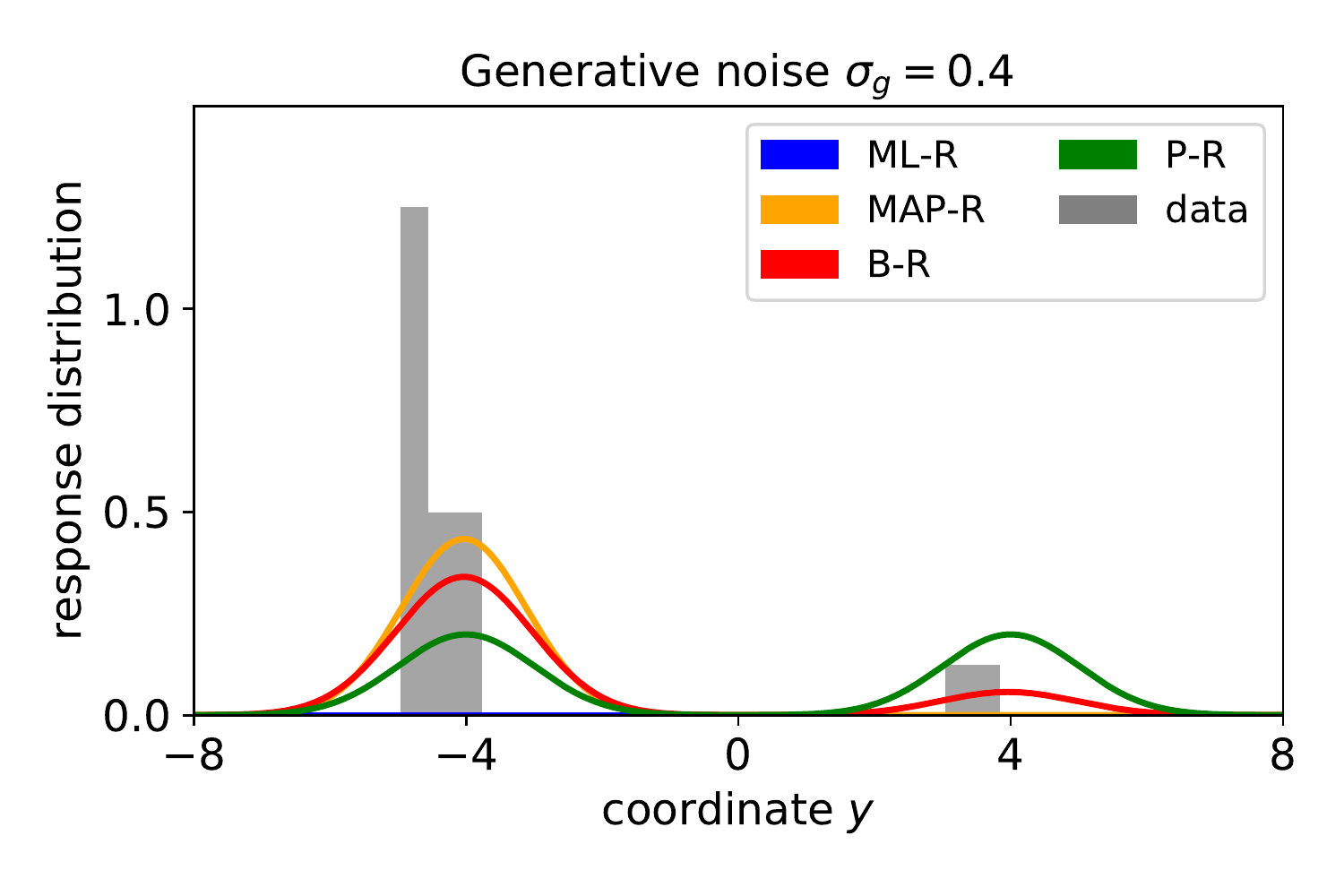}
\end{minipage}
\end{tabular}
\caption{ \label{fig:data-overview}
Example responses from a representative participant for different levels of generative noise $\sigma_g$. \textbf{(A)} A sample stimulus with the high noise level ($\sigma_g = 0.4$; blue dots), as shown to the participant. Contours indicate the equiprobable responses predicted by ML-R, MAP-R and B-R (not shown to the participant). The black dots show the measured responses. \textbf{(B-D)} The predicted response distribution (color-coded) vs. experimentally measured responses (gray) at $x^\star = 2$. As $\sigma_g$ increases, the data becomes less informative. Consequently, and in accordance with B-R, the response distribution becomes more bimodal. In the \textbf{(Bottom right)} plot, the mode of the M-L predictive distribution (blue) lays at high negative values, outside the plotting range.
}
\end{minipage}
\end{figure}

We considered four regression models. 
Maximum likelihood regression (ML-R) does not take the prior distribution into account and computes only the point estimate of $w$ that maximizes the likelihood $p(D_j|w)$. The maximum a posteriori model (MAP-R) combines the likelihood with the prior distribution to compute the mode of the posterior distribution $p(w|D_j)$. The Bayesian regression (B-R) model $\mathcal{M}_{BR}$ takes the entire posterior distribution into account:
\begin{align}
p(y^\star | x^\star, D_j, \mathcal{M}_{\text{BR}})  &= \int p(y^\star | x^\star, w) p(w|D_j) dw.
\label{eq:BR-prediction}
\end{align}
Note that in the previous three models, we assumed that participants know the generative noise ($\sigma_g$) -- an assumption that will be relaxed in \Cref{sec:var}.
As a null model, we included prior regression (P-R), which replaces the posterior with the prior, i.e., it does not use the likelihood. For more details, see Materials and Methods \Cref{sec:methods:inference-models}.
To model participants' responses, we also included the internal sources of noise arising during neural processing, decision making and the execution of motor action. To this end, we presented the same noise-free ($\sigma_g = 0$) stimulus 20 times and measured the variability of the responses $\sigma_m$. We then transformed the predictive distribution into a (predicted) response distribution over $r$ by convolving with a Gaussian $\mathcal{N}(r-y^\star;0,\sigma_m^2)$ (see Materials and Methods \Cref{sec:methods:internal-noise}). 
\Cref{fig:data-overview} shows the responses of a typical subject along with the model predictions. Both ML-R and MAP-R ignore one of the modes (here, the mode corresponding to a parabola which opens upwards). In addition, the parabola predicted by ML-R is much thinner (i.e., the absolute value of the quadratic parameter is high) than the parabola that the participant responded with because ML-R ignores the prior but humans likely do not. In the low noise regime ($\sigma_g=0.03$) (\Cref{fig:data-overview} \textbf{(B)}), the participant's unimodal response distribution rules out the null model, which shows a second mode. In the high noise regime ($\sigma_g=0.4$) (\Cref{fig:data-overview} \textbf{(D)}), MAP-R and ML-R fail to account for the fact that the participant distributed his responses across both modes. In the intermediate noise regime ($\sigma_g=0.1$) (\Cref{fig:data-overview} \textbf{(C)}), the participant sometimes gave tightly clustered responses and sometimes distributed them across both modes. Generally, B-R captures the participant's responses best as compared to the other models.

\subsection{B-R outperforms the other models}
\begin{figure}[pb]
\begin{minipage}{0.5\textwidth}
\centering
\begin{tabular}{ll}
{\bf (A)} & {\bf (B)} \\
\begin{minipage}{0.49\textwidth}
\includegraphics[width=\textwidth]{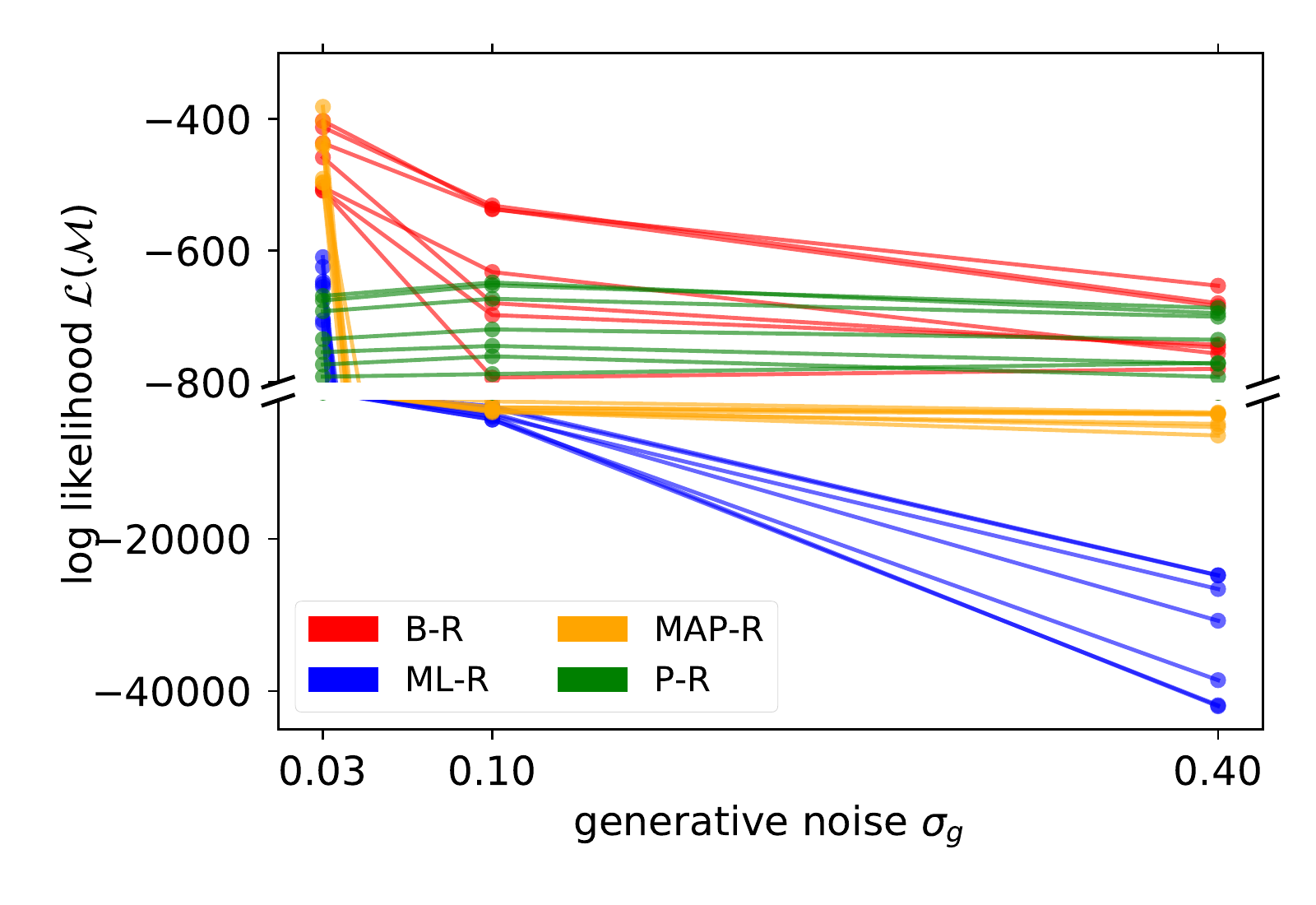}
\end{minipage} &
\begin{minipage}{0.49\textwidth}
\includegraphics[width=\textwidth]{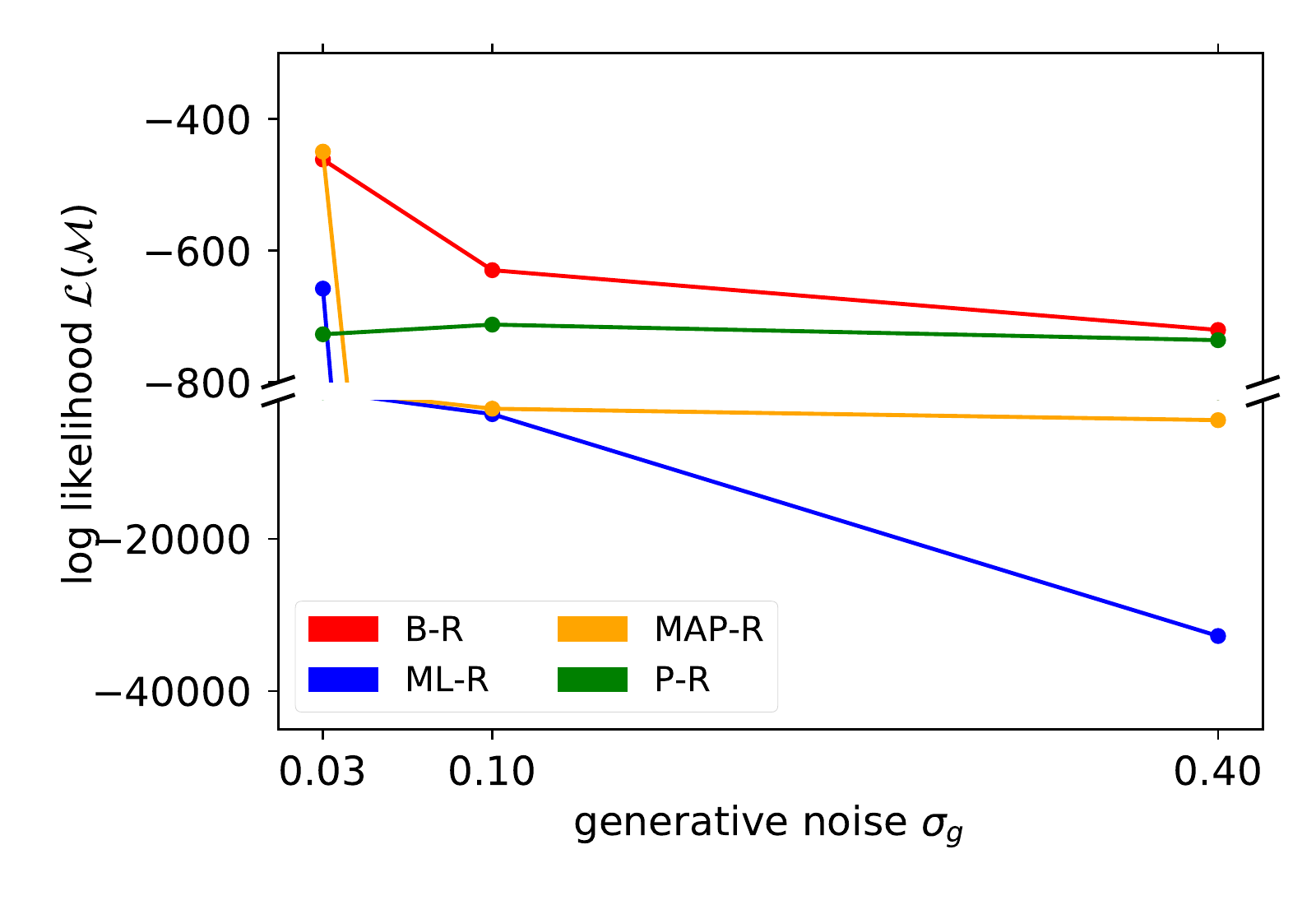}
\end{minipage}
\end{tabular}
\caption{ \label{fig:model-comparison}
The log-likelihoods \textbf{(A)} for all participants individually and \textbf{(B)} pooled across participants. The SEM is smaller than the markers. At $\sigma_g = 0.03$, ML-R and P-R show weak performance while MAP-R and B-R perform best and are indistinguishable within the errors. At $\sigma_g = 0.1$, the MAP-R performance drops relative to B-R. Finally, at $\sigma_g = 0.4$, B-R and P-R perform equally well while all other methods fail. The noise in the stimulus is so large that the posterior closely resembles the prior.}
\end{minipage}
\end{figure}

We showed the performance of a typical participant in \Cref{fig:data-overview}. Now, we compare model performance across all seven participants. For each of the seven participants individually, we computed the log probability that the model reproduces their responses, i.e., the y-position adjusted by the participant. We summed these log probabilities for all of the 20 stimuli $D_j$ as a measure of the quality of the model. \Cref{fig:model-comparison} shows these values for each noise level and participant separately (left) and averaged across participants (right). The higher the value, the better the model performance.

For all noise levels, B-R is among the highest performing models. At the lowest and highest noise level, MAP-R and P-R show similar performance to B-R, respectively. The results are consistent across participants. At low noise levels, MAP-R and B-R perform equally well because the stimulus is unambiguous. In this case, the posterior belief about the quadratic parameter is well approximated by a single value, the MAP. In contrast, at high noise, the prior is more important in determining the posterior belief about the quadratic parameter. In fact, the posterior closely approximates the bimodal prior. For this reason, B-R and P-R perform similarly well. 
At the medium noise level, the posterior belief is influenced by the prior and likelihood to approximately the same degree. In this case, B-R outperforms the other models because it takes full advantage of the posterior distribution, in particular, its bimodality. 
At medium and high noise levels, even P-R clearly outperforms MAP-R and ML-R because the latter two models predict unimodal responses, i.e., they predict that the parabola either opens up- or downwards depending on the stimulus $D_j$. 
Their failure to model responses belonging to the other mode is strongly punished by the log-likelihood measure because the model predicts a close-to-zero probability for these responses. 

\subsection{B-R explains the generative noise-dependent increase in response variance}
\label{sec:var}
When participants responded with a position that was far from a mode of the predicted response distribution, the log-likelihood strongly punished the corresponding model. To complement the log-likelihood analysis with a metric that is more sensitive to the overall response distribution, we used the variance. For each stimulus $D_j$, we computed the variance of the 20 recorded responses. For example, if all responses are located in the upper mode, the variance is close to zero. If they are distributed evenly across both modes, the variance is close to 16, which corresponds to the variance of the prior P-R. To obtain a single value for each participant and each noise level $\sigma_g$, we averaged the variance values across stimuli $j$. This value represents the averaged variance of responses at stimuli generated at $\sigma_g$. Next, we compared the average variance with the predicted averaged variance of each model. To this end, we generated a large number of stimuli $\{D_j\}_{j=1}^{1000}$ from $\sigma_g$ and averaged the variance of the predicted response distributions (see Materials and Methods \Cref{sec:methods:variance-comp}).

\begin{figure}[H]
\begin{minipage}{0.5\textwidth}
\begin{tabular}{ll}
{\bf (A)} & {\bf (B)} \\
\begin{minipage}{0.45\textwidth}
\includegraphics[width=\textwidth]{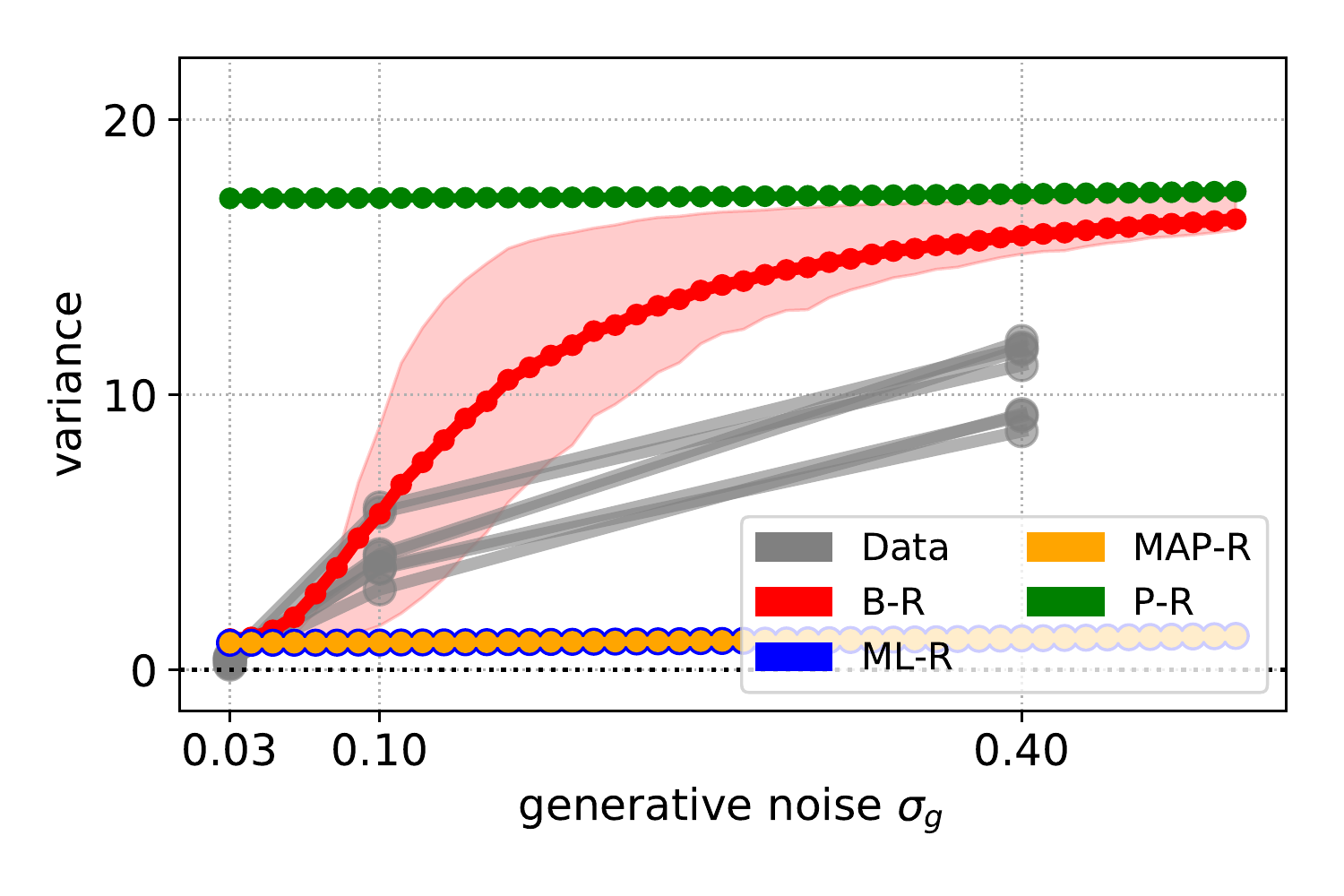}
\end{minipage} &
\begin{minipage}{0.45\textwidth}
\includegraphics[width=\textwidth]{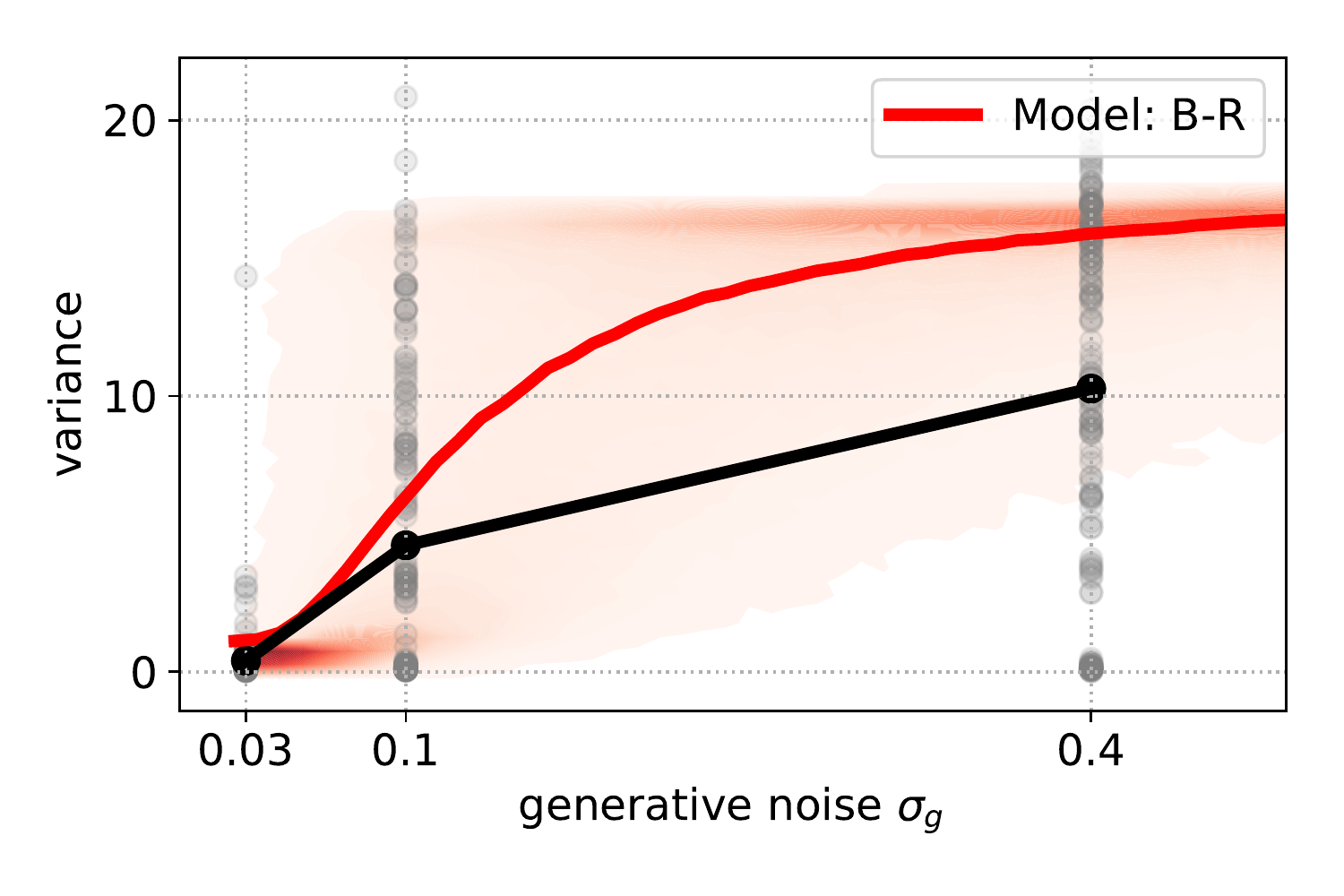}
\end{minipage} \\
{\bf (C)} & {\bf (D)} \\
\begin{minipage}{0.45\textwidth}
\includegraphics[width=\textwidth]{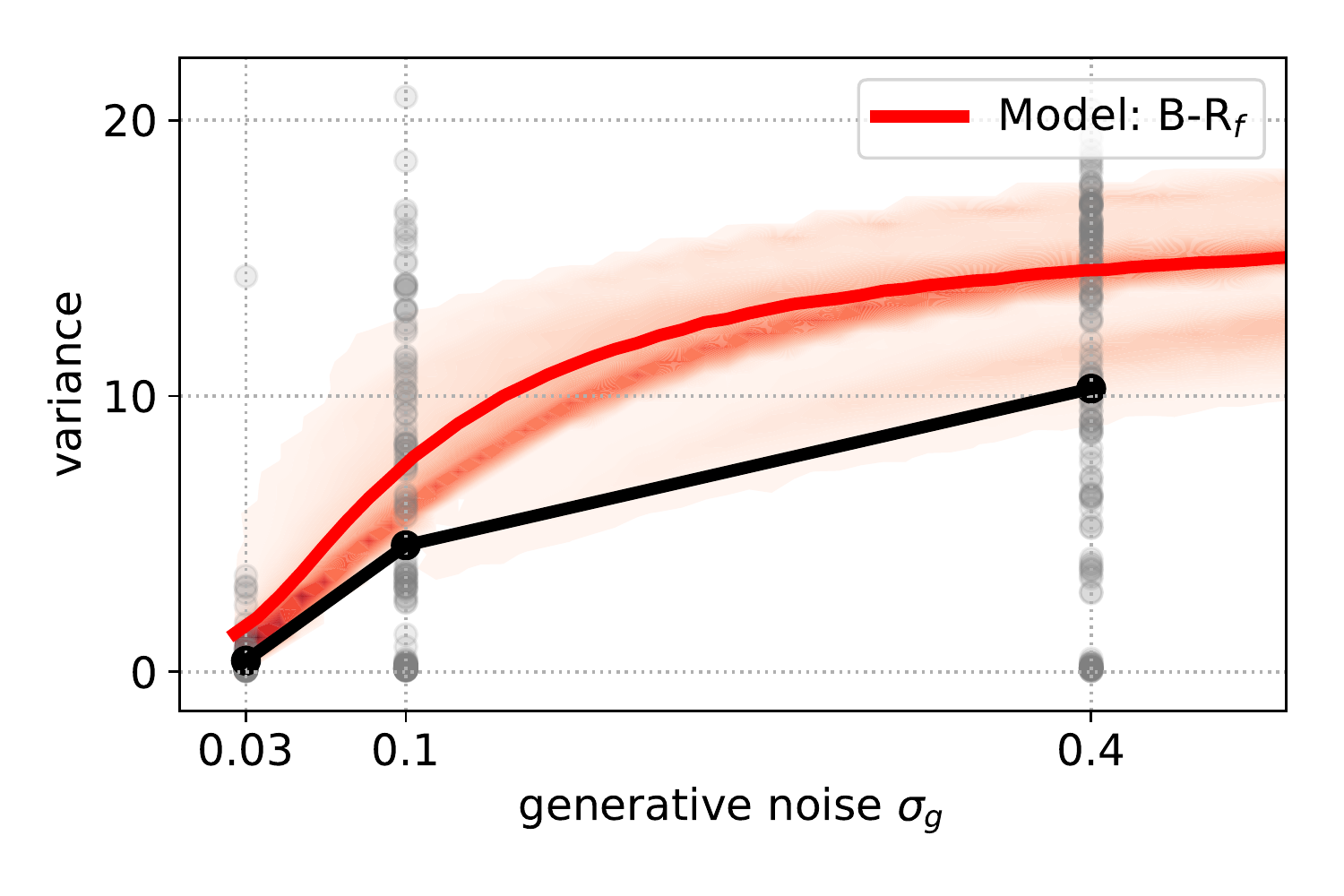}
\end{minipage} &
\begin{minipage}{0.45\textwidth}
\includegraphics[width=\textwidth]{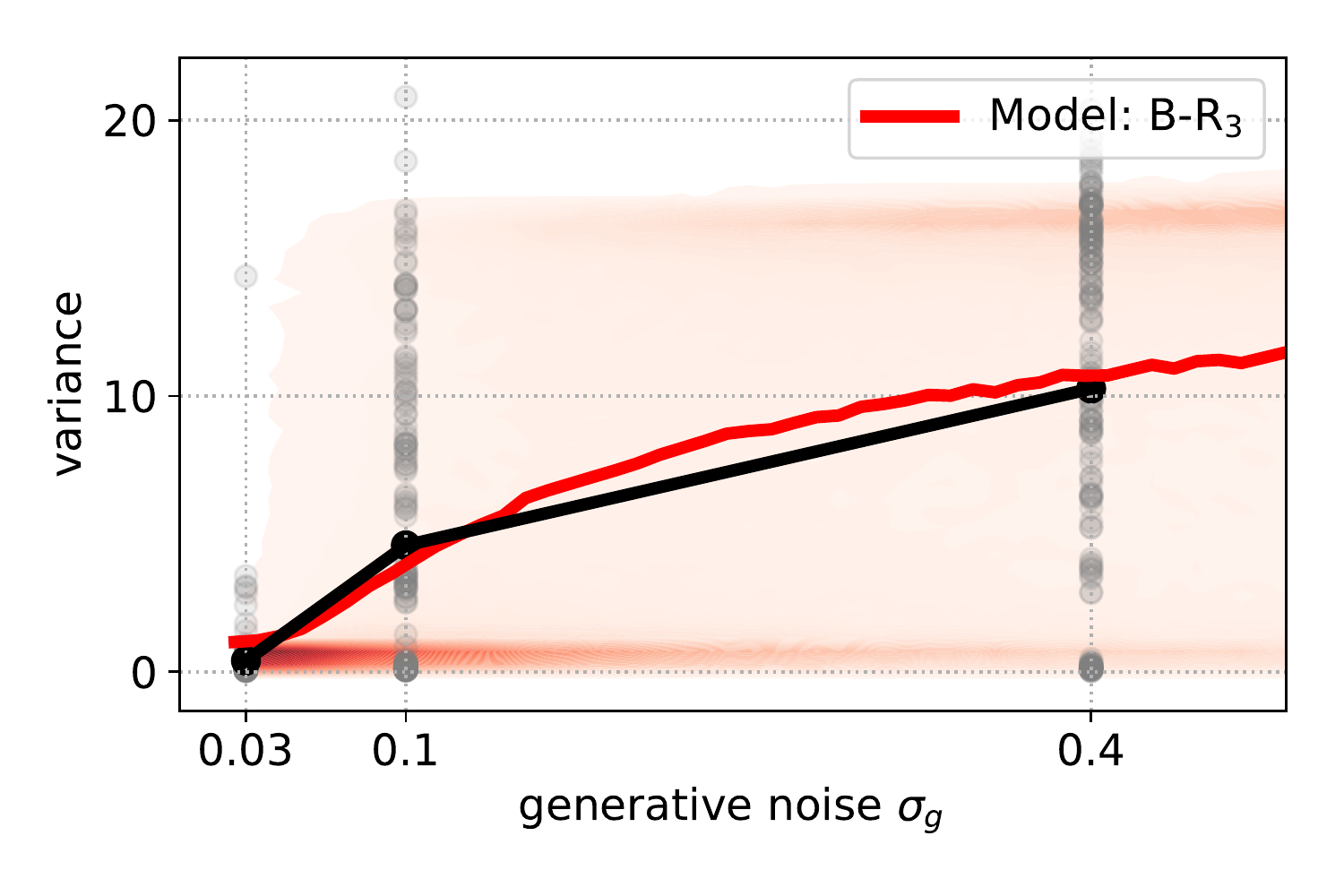}
\end{minipage}
\end{tabular}
\caption{ \label{fig:model-var-comparison}
\textbf{(
A)} Variance predicted by the models and averaged across participants (color-coded) and measured variance (gray) for individual participants.
B-R is the only model that captures the upwards trend in the data. The shading represents the 25\% and 75\% quantiles, computed from the distribution of variance across 1000 unique stimuli (and averaged over participants). While the errors are large enough to encompass all data points at $\sigma_g=0.1$, B-R fails at $\sigma_g=0.4$. The errors of the other models are too small to show.
\textbf{(
B,C,D)} Variance of the response distribution for a set of stimuli $\{D_j\}_j$, pooled across participants (gray dots, mean: black line) and predicted (red linear shading, mean: red line). Each of the 7 participants responded to 20 stimuli, yielding 120 gray points per generative noise $\sigma_g$. Some stimuli always elicit an almost zero variance. As $\sigma_g$ increases the distributions of variances moves upwards. The large spread of variances shows that participants distribute individual responses anywhere from very clustered in one of the modes to equally spread out across both modes, depending on the specific stimulus. We simulated 1000 stimuli per $\sigma_g$ to obtain the predicted variance distribution. \textbf{(B)} Standard B-R captures the upwards trend in the data but fails to account for small variance at $\sigma_g = 0.4$. \textbf{(C)} B-R$_f$ cannot account for the almost zero variances. However, as $\sigma_g$ increases it predicts that the response variance fall into two clusters, i.e., a second lobe appears below the mean. Likewise, the gray dots seem non-homogeneously distributed but we do not have enough data to quantitatively test this. \textbf{(D)} B-R$_3$ accounts for the the almost zero variance because fitting $\sigma_g$ on each trial occasionally yields very high certainty and unimodal responses. It also fits the mean responses well. 
}
\end{minipage}
\end{figure}

\Cref{fig:model-var-comparison} (top left) compares the averaged variance of the participants' responses (gray) with the model predictions (color-coded). 
The B-R variance changes strongly with the noise level. The reason is that the generative noise $\sigma_g$ determines the contribution of likelihood and prior for the responses; thus, the predicted responses change from a unimodal to bimodal in B-R. As a result, the B-R (red) variance values smoothly transition from the MAP/ML variances at the low noise level ($\sigma_g = 0.03$) to the P-R variance at the high noise level ($\sigma_g = 0.4$). This is not the case in the other models.

There is some discrepancy between the data from the participants and the prediction of B-R. At high noise ($\sigma_g=0.4$), the participants' variance increases more slowly than what B-R predicts. This means that participants cluster their responses close to one of the modes more often than predicted. 

To explain the discrepancy, we tested two additional variants of B-R. Full Bayesian regression B-R$_f$ and three-point Bayesian regression B-R$_3$. Both are based on the idea that participants do not regard the generative noise as constant but rather infer (in B-R$_f$) or estimate (in B-R$_3$) it on a per-trial basis. When participants assumed that the generative noise was smaller than the one we used to generate the points, their responses were more clustered. As a consequence, the average variance decreased. This happened when the four stimulus points were by chance aligned on a parabola. When we computed the averaged variance with classical B-R, we used the true generative noise which was, in this case, larger than the estimated noise. Since the noise determines how much participants rely on the prior and likelihood relative to each other, the averaged variance of the prediction was larger than in the data.

Both additional variants of B-R consider the noise on a trial-by-trial basis, but they differ in how they do it. While B-R only infers the quadratic parameter $w$, B-R$_f$ extends the inference to the noise level $\sigma_g$ as well. 
Since the generative model does not specify a prior over the (fixed) hyperparameter noise $\sigma_g$, we used an exponential prior and set its mean to the fixed hyperparameter (see Material and Methods \cref{eq:BRf-prediction}). B-R$_f$ uses the joint posterior over $w$ and $\sigma_g$ to predict further responses. When the 4-dot stimulus closely resembled a parabola, the posterior belief over $\sigma_g$ was highly peaked around a value that was smaller than the actual generative noise parameter. Thus, we expected more clustered responses and predicted a lower averaged variance.

B-R$_3$ differs from B-R in that it includes the preprocessing step of estimating the generative noise on a trial-by-trial basis and treating the left or rightmost stimulus point as an outlier. The noise estimate is then used in standard B-R. The easiest approach to achieve this is using the maximum-likelihood estimator of the noise and all four stimulus points. Often, the maximum likelihood estimate was smaller than the fixed hyperparameter, reducing the averaged variance (see Supplementary Information). Data and prediction matched almost perfectly if we allowed for the possibility to disregard outliers during the estimation (see Materials and Methods \cref{sec:methods:B-R3}). 

\Cref{fig:model-var-comparison} (top right and bottom) show the variance distribution predicted (red, mean: line) by B-R, B-R$_f$ and B-R$_3$. The data (gray, mean: black) is pooled across participants. In the data, there is a portion of almost zero-variance responses for all noise levels. The fraction of these highly clustered responses decreases as the generative noise increases but remains present for all experimental conditions. Only B-R$_3$ captures this aspect of the data. Besides this, the variances associated with both variants of B-R rise more slowly than that of the original B-R, supporting the idea that the generative noise is indeed jointly estimated with the quadratic parameter.
The comparison between measured and predicted responses strongly favours B-R. B-R with $\sigma_g$ as a fixed hyperparameter overestimates the increase in variance compared to the data. By using two variants of B-R, we show that this discrepancy can be explained by assuming that participants estimate the generative noise on a per-trial basis. The two additional variants of B-R perform slightly worse than the original B-R in terms of the log-likelihood but still better than all other models (see Supplementary Information).

In summary, both the likelihood-based model comparison and the response variance results strongly favour B-R.


\section{Discussion}

Participants adjusted a dot such that it fits on a parabola determined by four other dots. We used the variance 
and the log-likelihood to compare the participants' responses to the predictions of four regression models: ML-R, MAP-R, P-R and B-R. Out of these, B-R best explained the participants' responses across different levels of stimulus noise. In particular, B-R interpolated between the variances 
of MAP-R at low noise levels and P-R at high noise levels. We found the same trend in the data, although B-R predicted larger values for the variance 
at high noise levels. We could explain this discrepancy with the assumption that participants estimate the noise level per trial and then apply it. We conclude that in our task, participants are capable of implicitly performing or at least approximating a complicated integral over the posterior distribution of the parameter given the data.

The rationale behind our experimental design was to use the simplest possible setup to study if humans perform Bayesian regression. Our one-dimensional response space was easy to visualize and the amount of data needed to compare the predictive and empirical distributions was limited. 
Bayesian regression easily extends beyond the simple low dimension problem of a single quadric nonlinearity to problems with higher dimensional parameters, such as polynomials of arbitrary degree, and to other basis functions, such as Gaussians. Future research needs to study to what level of complexity humans perform Bayesian regression. There is some evidence that the brain uses sampling to represent high dimensional distributions \citep{fiser2010statistically}. This representation of distributions scales to high dimensions and offers a potential way of evaluating the high dimensional integral needed for Bayesian regression.

In our analysis, we assumed that participants know the generative model, including the prior over the parameters. Without this assumption, we would have had to account for potential temporal dynamics of learning with a participant-specific, time-dependent prior. For instance, it might take participants a non-negligible amount of time to learn the generative model or their responses could be influenced by immediately preceding trials. To avoid such complications, we showed the generative parabola after each trial.

Our work was inspired by the growing emphasis on parameter uncertainty in the machine learning community; however, it is important to highlight that function learning and extrapolation have been studied before. The function learning literature has addressed which types of functions humans can learn \citep{koh1991function}, how batch or sequential data representation affects learning \citep{villagradata2018}, to what extent human behaviour can be modelled by parametric functions \citep{mcdaniel2005conceptual} and how well humans extrapolate \citep{delosh1997extrapolation}. However, to the best our knowledge, these studies have so far failed to conduct a minimal experiment to establish that humans behave in accordance with Bayesian regression. Our contribution will help to better understand the brain’s remarkable ability to learn and generalise from very little data and underpins the power of Bayesian regression as a framework in psychophysical modelling.


\matmethods{
\label{sec:methods}
\subsection{Stimulus generation from the bimodal prior}
\label{sec:methods:stimulus-generation}
Here, we describe in detail how stimuli are generated. On the j$^{th}$ trial, participants are presented with a stimulus consisting of $N=4$ points in a 2-dimensional space: $D_j = \{ (x_1^{(j)} = -0.3,y_1^{(j)}),(x_2^{(j)} =-0.1,y_2^{(j)}),(x_3^{(j)} =0.1,y_3^{(j)}),(x_4^{(j)} =0.3,y_4^{(j)}) \}$. For each stimulus, we fix the x-values and generate the y-coordinates from a Gaussian generative model with a parabolic non-linearity and the \textit{generative parameter}, $w_j$:
\begin{align}
\label{eq:generating_parabola}
p(y|x, w_j) &= \mathcal{N} (y;w_j x^2,\sigma_g^2).
\end{align}
The parameter $w_j$ is drawn from a mixed Gaussian prior
\begin{align}
\label{eq:prior}
\pi_\gamma(w_j) = \left( c\mathcal{N}(w_j;\mu_\pi,\sigma_\pi^2) + (1-c) \mathcal{N}(w_j;-\mu_\pi,\sigma_\pi^2)\right)
\end{align}
where the parameter set $\gamma = (\mu_\pi, \sigma_\pi^2,c)$ consists of the mean $\mu_\pi = 1$, the standard deviation $\sigma_\pi = 0.1$ and the mixing coefficient $c = 1/2$. We denote the total set of hyperparameters (suppressed for notational clarity), from the prior and the generative probability, by $\alpha = (\gamma,\sigma_g^2)$. Each parameter $w_j$ corresponds to a \textit{generative parabola}. Given this model and given a stimulus $D_j$, we asked participants to predict the y-component $y^\star$ at $x^\star = 2$, which is equivalent to mentally fitting a parabola to the four stimulus points and estimating the point of intersection with a vertical line at $x^\star$.\\
To train participants on the generative model and the prior, we showed participants the generative parabola after each trial. 
In total, we showed a set of 20 unique stimuli for each of the three noise levels $\sigma_g \in \{0.03,0.1,0.4\}$, and each unique stimulus was repeated 20 times. We denote the set of the 20 responses to the j$^{th}$ stimulus as $R_j = \{r^{(j)}_1, \dots r^{(j)}_{20}\}$. This amounts to a total of 400 trials per noise level. The order of the stimuli was randomized. \Cref{fig:psychophysics-setup} shows the set-up.

\subsection{Regression models}
\label{sec:methods:inference-models}
In each trial, the participants carried out an inference step and a prediction step. During the inference step, they inferred information about the quadratic parameter $w_j$ based on the presented data (i.e., stimulus) $D_j$. The information they inferred depends on the inference model participants use. The inferred information was then used for a subsequent prediction $y^\star$. Therefore, the participants overall task was to compute the \textit{predictive distribution}: $p(y^\star | x^\star, D_j, \mathcal{M})$. \\
\textit{\textbf{Prior regression}} (P-R) is our null model. P-R assumes that participants make predictions based on their prior belief but disregard information from the stimulus:
\begin{align}
\label{eq:Prior-prediction}
p(y^\star | x^\star, D_j, \mathcal{M}_{PR})  &= \int p(y^\star | x^\star, w) \pi(w) dw
\end{align}
\textit{\textbf{Maximum likelihood regression}} (ML-R) relies only on the likelihood maximizing parameter, $w_{ML}$:
\begin{align}
p(y^\star | x^\star, D_j, \mathcal{M}_{\text{ML}}) &=\notag p(y^\star | x^\star, w_{ML}) \\ 
& \text{with} \ \ w_{ML} = \arg \max_{w} p(D_j | w)
\label{eq:ML-prediction}
\end{align}
\textit{\textbf{Maximum a posteriori regression}} (MAP-R) uses the parameter that maximizes the posterior $p(w|D_j)= p(D_j | w)\pi(w)/ p(D_j)$:
\begin{align}
p(y^\star | & x^\star, D_j, \mathcal{M}_{\text{MAP}}) = \notag p(y^\star | x^\star, w_{MAP}) \\ 
& \text{with} \ \ w_{MAP} = \arg \max_{w} p(w|D_j).
\label{eq:MAP-prediction}
\end{align}
\textit{\textbf{Bayesian regression}} (B-R) uses the entire posterior for making predictions by marginalizing over it:
\begin{align}
p(y^\star | & x^\star, D_j, \mathcal{M}_{\text{BR}})  = \int \notag p(y^\star | x^\star, w) p(w|D_j) dw \\ & \text{with} \ \ p(w|D_j) = p(D_j | w)\pi(w) p^{-1}(D_j)
\label{eq:BR-prediction}
\end{align}
\label{sec:methods:B-Rf}
\textit{\textbf{Full Bayesian regression}} (B-R$_f$) loosens the assumption that participants treat $\sigma_g$ as a hyperparameter. Including $\sigma_g$ with an exponential prior $\pi(\sigma_g)$ in Bayesian regression generalizes \cref{eq:BR-prediction}:
\begin{align}
p(y^\star | x^\star, & D_j, \mathcal{M}_{\text{BR}_f}) = \int p(y^\star | x^\star, w, \sigma_g')  p(w, \sigma_g'|D_j) \text{d}w \text{d}\sigma_g' \notag 
\\ 
& \text{with} \ \ p(w,\sigma_g'|D_j) = p(D_j | w, \sigma_g')\pi(w) \pi(\sigma_g') p^{-1}(D_j).
\label{eq:BRf-prediction}
\end{align}
We chose an exponential prior because it has only one free parameter. By setting the expectation over $\sigma_g'$ equal to the actual hyperparameter $\sigma_g$, we eliminated this degree of freedom and retained our fitting free approach (see Supplementary Information).
\label{sec:methods:B-R3}
\textit{\textbf{Three point regression}} (B-R$_3$) assumes that participants use an estimate $\hat{\sigma}_g$ based on $M=3$ points. The posterior predictive is computed exactly as in ordinary B-R (\cref{eq:BR-prediction}) except for the following replacement ${\sigma}_g \rightarrow \hat{\sigma}_g,$ where $\hat{\sigma}_g$ is computed in a two step procedure. First, we obtained two maximum likelihood estimates of the generative noise by ignoring the left most and the right most point. Second, the minimum of these two was $\hat{\sigma}_g$. We also tested $M=4$ as an alternative but concluded that $M=3$ gave a more convincing account of the observed variance (see Supplementary Information Figure S1).\\
\subsection{Participants' internal noise}
\label{sec:methods:internal-noise}
To predict the participants' responses $r$ from the regression models' output $y^\star$, we had to account for the internal noise of the participants. We did this by showing a noise-free parabola and fitting a Gaussian with variance $\sigma_m^2$ to each participant's response variability: $p(r|y^\star) = \mathcal{N}(r; y^\star, \sigma_m^2)$. As explained in more detail in the Supplementary Information, the predicted response distribution is then
\begin{align}
p(r | D_j, x^\star, \mathcal{M}) = \int p(r | y^\star) p( y^\star | x^\star, D_j, \mathcal{M}) dy^\star.
\label{eq:predict-observers-single-r}
\end{align}
\subsection{Model comparison} We used the log-likelihood and the variance to compare the predicted and empirical response distributions.\\
\textit{\textbf{Log likelihood.}}
To compute the log likelihood for a model $\mathcal{M}$ across all response at a given noise level $\sigma_g$, we summed the individual log likelihoods of each response $r$ (the log of \cref{eq:predict-observers-single-r}) across all stimuli $D_j$:
\begin{align}
\mathcal{L}_\mathcal{M} := \sum_{j=1}^{20} \sum_{r \in R_j} \log p(r | D_j, x^\star, \mathcal{M}).
\label{eq:predict-observers-behavior2}
\end{align}
\textit{\textbf{Variance prediction}}
\label{sec:methods:variance-comp}
As a independent comparison of the data and the predicted response distribution, we used the variance. For each of the 20 stimuli $D_j$ we obtained a single empirical value from the 20 responses recorded. Hence, at each noise level $\sigma_g$, we have a variance distribution:
\begin{align}
V_{exp} = \left\{ \tfrac{1}{19} \sum_{k=1}^{20} (\bar{r}^{(j)} -r^{(j)}_k)^2
| j \in \{1, \dots, 20 \},
\right\}
\end{align}
where high variance values reflect ambiguous and difficult stimuli while low values indicate easy stimuli, prompting participants to give very similar responses across trials.
The predicted variance distribution is expressed by
\begin{align}
V_{theo} = \left\{ \text{Var}[r | D_j, x^\star, \mathcal{M}]
| j \in \{1, \dots, 1000 \}
\right\},
\end{align}
where we used 1000 stimulus samples and computed the variance analytically (see Supplementary Information). We use this distribution to compute the means and quantiles shown in \cref{fig:model-var-comparison} (top left) and the color-coded density in the other plots.

\subsection{Participants}
\label{sec:methods:participants}
Seven participants (3 females, 4 males, ages 21-27) participated in the experiment. The experiment was programmed using custom software implemented in MATLAB. Stimuli were presented on a 1920x1080 (36 pixels/cm) monitor with a refresh rate of 120 Hz. Participants viewed the display binocularly. Each trial comprised a fixation dot presented for 1 s followed immediately by presentation of the stimulus (with 5 arcmin point diameter). Participants moved a red point up or down using the up and down arrow keys to indicate the vertical position of the parabola at the given horizontal location. See Supplementary Information for more details.
}

\showmatmethods{} 

\acknow{We would like to thank Frank Jäkel for his insightful input. J.J. and J.-P.P. were supported by the Swiss National Science Foundation grants PP00P3\_150637 and 31003A\_175644. M.A.J., M.P. and M.H.H. were supported by the SNF grant 'Basics of visual processing: from elements to figures' (176153).}

\showacknow{} 

\bibliography{pnas-sample}

\end{document}



\maketitle

\SItext
\section{Experiment}
\subsection{Participants}
\label{sec:methods:participants}
Seven participants (3 females, 4 males, age 21-27) participated in the experiment. All participants had normal or corrected-to-normal visual acuity, as measured with the Freiburg Visual Acuity Test (Bach, 1996). Participants were paid 20 CHF/hour. All participants gave informed consent in accordance with protocol 384/2011 “Commission cantonale d’éthique de la recherche sur l’être humain”. Participants provided written consent prior to the experiment.

\subsection{Apparatus and Stimuli}
\label{sec:methods:apparatus}
The experiment was programmed using custom software implemented in MATLAB with the Psychophysics Toolbox (Brainard, 1997). Stimuli were presented on a gamma-corrected ASUS VG248QE LCD monitor with a resolution of 1920x1080 (36 pixels/cm) and refresh rate of 120 Hz. Stimuli consisted of four black points (5-arcmin diameter) sampled from the generating parabola described and jittered vertically with Gaussian noise of standard deviation. On each trial, the width of the generating parabola was sampled from the prior distribution (as described in the methods section of the main text).
\\
Participants viewed the display binocularly, and a chin rest stabilized viewing distance at 75 cm from the screen. The origin of the Cartesian coordinate system (50 pixels/unit), according to which the positions of the stimulus points were defined, was placed at the center of the screen. Each trial comprised a fixation dot (5 arcmin diameter) presented at the center of the screen for 1 s followed immediately by presentation of the stimulus. Along with the four stimulus points, a red point of the same size was presented 2 units to the right of the screen center and participants were instructed to adjust this point up or down using the arrow keys to indicate the y-axis location at which the generating parabola would pass. Participants were given unlimited time to make their response.\\
\\
In order to allow for learning of the prior distribution of the quadratic parameter, feedback was provided after each trial by overlaying on the stimulus points the generating parabola with a green line following each response. A 1 s delay (with fixation) preceded the next trial.\\

\subsection{Experimental Design}
\label{sec:methods:experimental-design}
The experiment was divided into two sessions, completed on two separate days. Each session began with noiseless trials ($\sigma_g = 10^{-5}$), to familiarize participants with the paradigm and, on the modeling side, to allow fitting of the participants' motor noise $\sigma_m$. Following these initial practice trials, each session consisted of 12 blocks of 50 trials each. The first 4 in both of the sessions comprised stimulus points with low noise ($\sigma_g = 0.03$). participants then completed 8 blocks of medium to high noise trials, with $\sigma_g = 0.1$ in session 1 and $\sigma_g = 0.4$ in session 2. \\
\\
In total, we showed a set of 20 unique stimuli for each of the three noise levels $\sigma_g \in \{0.03,0.1,0.4\}$, and each unique stimulus was repeated 20 times. We denote the set of the 20 responses to the j$^{th}$ stimulus as $R_j = \{r^{(j)}_1, \dots r^{(j)}_{20}\}$. This amounts to a total of 400 trials. The order of the stimuli was randomized for each participant.

\section{Fitting the noise}
\begin{figure}[H]
\begin{minipage}{1.0\textwidth}
\begin{tabular}{ll}
{\bf (A)} & {\bf (B)} \\
\begin{minipage}{0.45\textwidth}
\includegraphics[width=\textwidth]{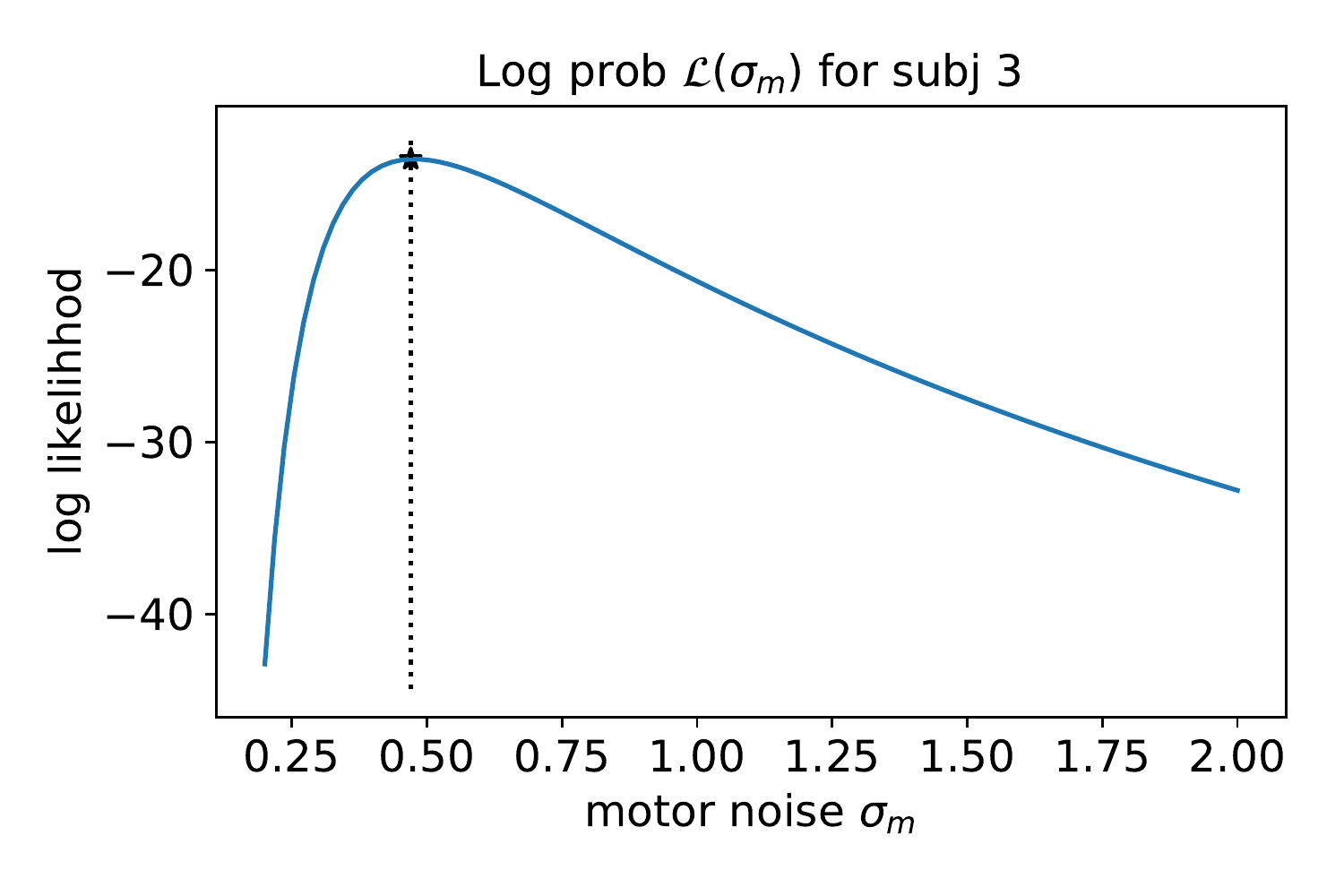}
\end{minipage} &
\begin{minipage}{0.45\textwidth}
\includegraphics[width=\textwidth]{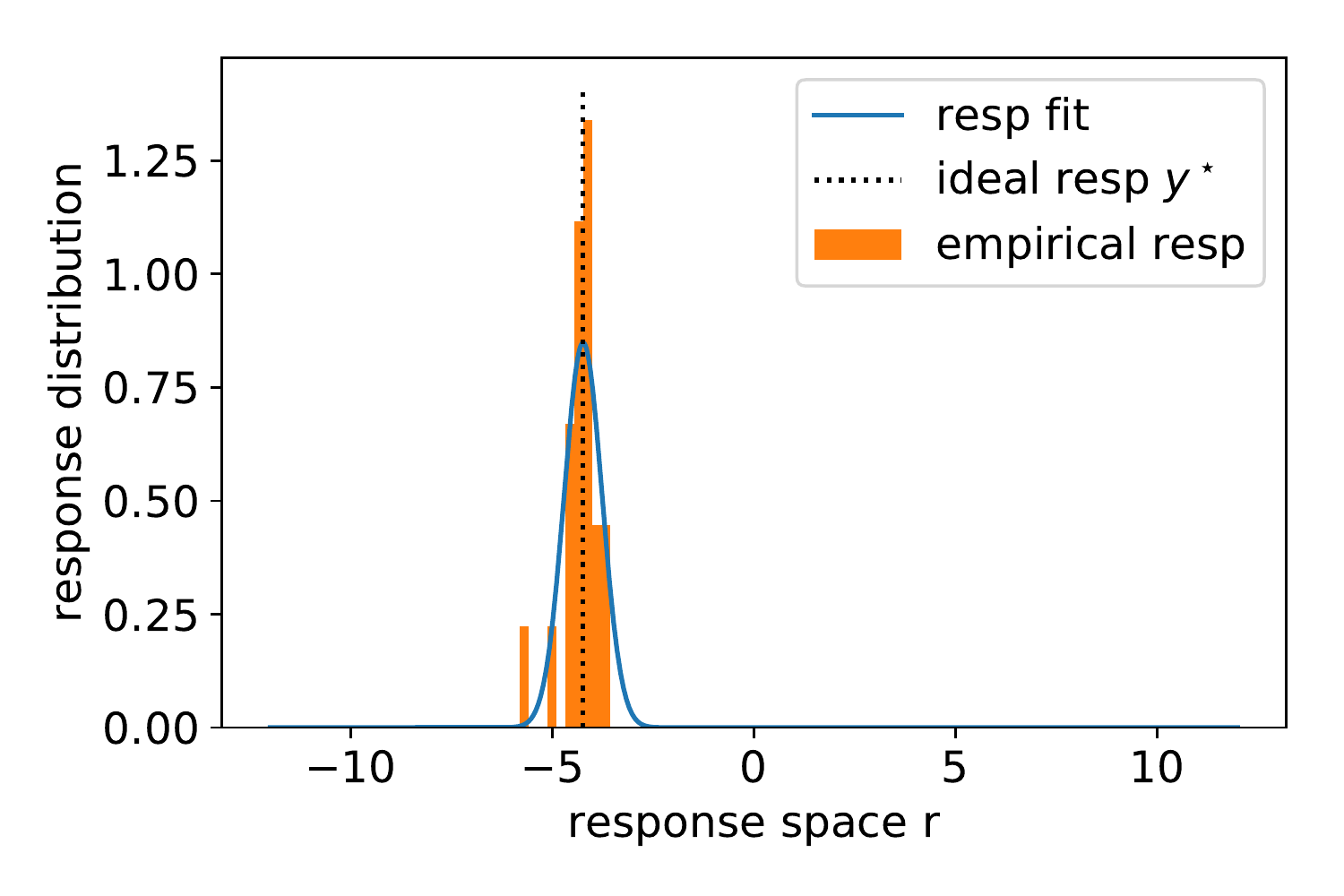}
\end{minipage}
\end{tabular}
\caption{\label{fig:pt-response-fct} \textbf{(A)}: The 3$^{rd}$ participant's log-likelihood landscape for $\sigma_m$. \textbf{(B)} Measured responses (orange histogram) of the participant and the response function (blue).}
\end{minipage}
\end{figure}

To quantify the internal noise sources of the participants, we measure the variability of their responses in the absence of parameter uncertainty. To eliminate the parameter uncertainty we run 20 noiseless trials. For the noiseless trails we generate a single stimulus $D_0$ with minimal\footnote{Our analysis code relies heavily on the precision $\sigma_g^{-2}$. To avoid the singularity at $\sigma_g=0$, we ran with $\sigma_g = 10^{-5}$ instead. Within the limits of perception in our setup, the resulting stimuli and the fit of the internal noise are not affected by this.} generative noise $\sigma_g = 10^{-5}$. In this case, the likelihood dominates the predictive distribution for all models, except for the null model which is independent of the likelihood. Because there is no parameter uncertainty left, all models predict the same sharply peaked answer $y^\star_0$. More formally, the data dependent factor used for computing the predictive distribution becomes a delta-function around $y_0^\star$ (see \cref{sec:noise-limits} for the derivation):
\begin{align}
p(y^\star | x^\star, D_0, \mathcal{M}) = \delta(y^\star - y_0^\star)
\end{align}
In the noiseless setting, the response variability can be fully attributed to internal noise sources such as limits in the perception of the location of the response, a noisy internal representation of the desired response and/or the inability to carry out precise motor commands. For the sake of simplicity, we will call \emph{motor noise} the joint effect of those factors while remembering that it summarizes several factors. We model this motor noise as a Gaussian such that the response distribution is given by
\begin{align}
p(r|y^\star) &= \mathcal{N}(r; y^\star, \sigma_m^2)
\label{eq:point-response-fct}
\end{align}
where $\sigma_m^2$ is the variance of the motor noise. Fitting $\sigma_m^2$ amounts to computing the variance across the responses when the participant is presented with $D_0$. \Cref{fig:pt-response-fct} (A) shows the log-likelihood fit of $\sigma_m$ for participant 3 and (B) the empirical response distribution versus the fitted point response function. Other participants behave correspondingly, except occasional outliers. These are most likely due to the fact that participants were still familiarizing themselves with the experimental design.

\section{Log likelihood for Bayesian Regression variants}
\begin{figure}[H]
\begin{minipage}{1.0\textwidth}
\begin{tabular}{ll}
{\bf (A)} & {\bf (B)} \\
\begin{minipage}{0.49\textwidth}
\includegraphics[width=\textwidth]{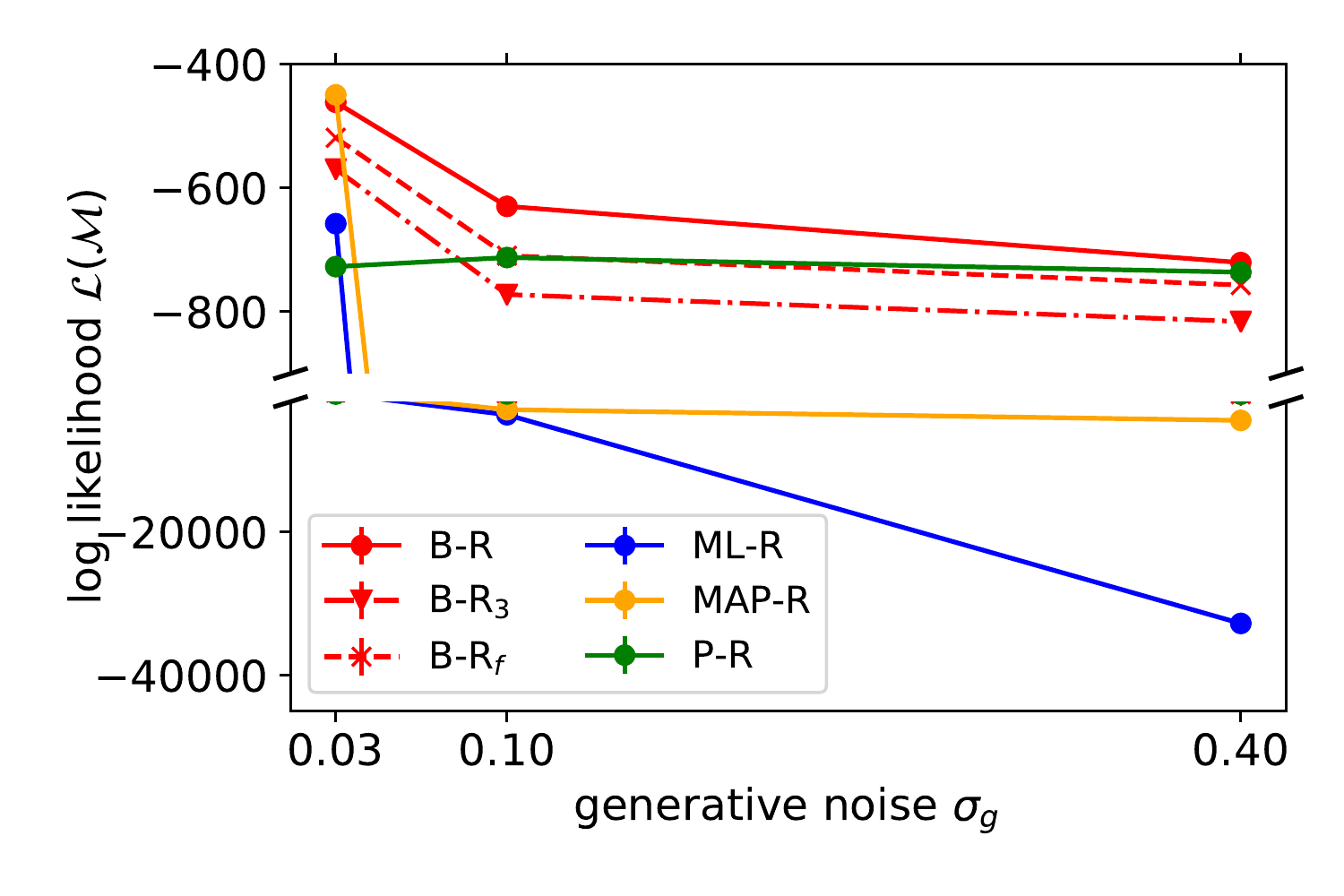}
\end{minipage}
&
\begin{minipage}{0.49\textwidth}
\includegraphics[width=\textwidth]{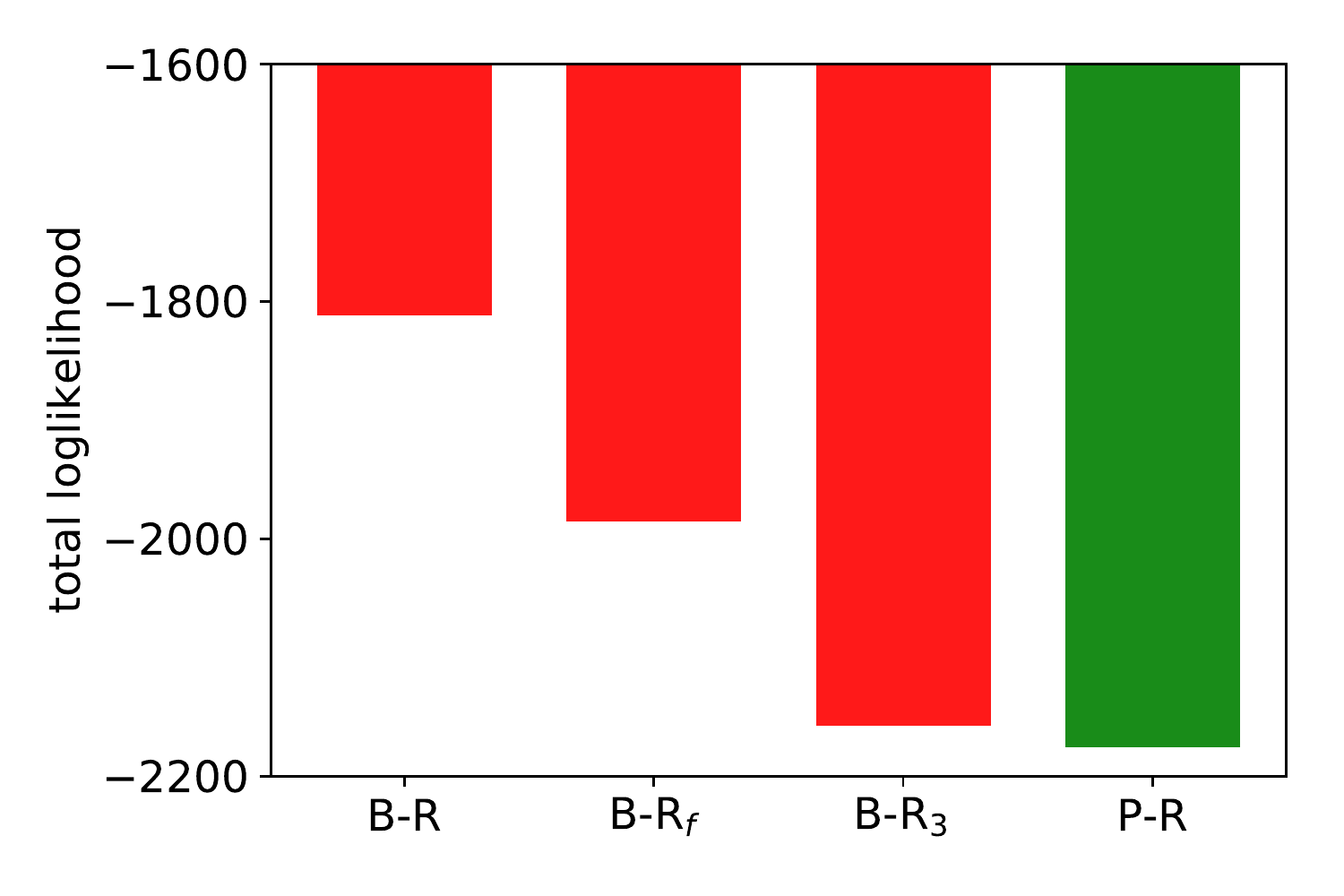}
\end{minipage}
\end{tabular}
\caption{\label{fig:all-log-like} Loglikelihood for all models. A poor fit to the data yields more negative loglikelihood; a good fit results in less negative values.  \textbf{(A)} Variants of B-R (red) perform less well than original B-R. \textbf{(B)} The total loglikelihood summed over the three levels of generative noise $\sigma_g$ and averaged across participants. ML-R and MAP-R are not shown because their poor performance is visually evident from plot (A).}
\end{minipage}
\end{figure}

In the main text, we introduced two additional variants of Bayesian regression (B-R): B-R$_3$ and full B-R$_f$. \Cref{fig:all-log-like} \textbf{(A)} displays the log-likelihood of all variants of B-R (red), the prior regression (green) as well as ML-R (blue) and MAP-R (yellow).
Out of the three B-R variants, the classical version performs best, then B-R$_f$ and finally B-R$_3$. At $\sigma_g=0.1$ and $\sigma_g=0.4$, B-R$_3$ describes the data worse than the prior. However, when we compute the loglikelihood for the entire data set by summing over $\sigma_g$, B-R$_3$ still slightly outperforms the prior, as shwon in \Cref{fig:all-log-like} \textbf{(B)}. 
Note that the additional B-R variants perform better than the original B-R in terms of the variance metric (see main text) but worse in terms of the likelihood. The reason is that the loglikelihood punishes responses far from the prediction strongly while the variance measure is less sensitive to wrong answers but measures the clustering of responses instead. Based on their trial-by-trial estimation, the additional variants of B-R assume a smaller level of generative noise. This leads to more confident predictions. Hence, when a response does not align with the prediction, it is punished more strongly than in the case of classical B-R.

\section{Derivations of predictive distributions}
\label{sec:supp:ppds-overview} 
Section B of the Materials and Methods section of the main text formally expresses the inference and prediction steps based on Maximum Likelihood regression (ML-R), Maximum a Posteriori regression (MAP-R), Bayesian Regression (B-R) and prior-regression (P-R), as well as the two variants of B-R, full Bayesian regression (B-R$_f$) and three point B-R (B-R$_3$). In order to turn these formal expressions into explicit functions of the parameters, a couple of algebraic steps are required. In the following we report these steps for a more general case than the one presented in the main text. Specifically, we generalize the one-dimensional curvature parameter $w$ to arbitrary dimensions $p$, consider any real nonlinear function $\boldsymbol{\phi}(x)$ rather than a simple quadratic and a mixed Gaussian prior without the symmetry imposed in the main text.

\subsection{Notation}
First, we highlight the notation used during the derivation as compared to the main text in some detail. The fact that the parameter and nonlinear function are now vector valued is indicated with bold symbols:
\begin{align}
w \rightarrow \textbf{w} \in \mathbb{R}^{p} \hspace{2cm} \phi \rightarrow \boldsymbol{\phi}: \mathbb{R} \rightarrow \mathbb{R}^{p},
\end{align}
For example, a polynomial of order $p-1$ can be expressed by $y(x)=\textbf{w}^T \boldsymbol{\phi}(x)$ by setting $\boldsymbol{\phi}(x) = \sum_{k=0}^{p-1} x^k \hat{e}_k$, where $\hat{e}_k$ is a unit vector.\\
In the main text, we denote the j$^{th}$ stimulus by $D_j = (\textbf{x}^{(j)},\textbf{y}^{(j)})$. Here, the bold notation indicated that we represented the $x$ and $y$ components of the $N=4$ stimulus points as a $N$ dimensional vector. In what follows, it is unnecessary to distinguish between different stimuli; however, it will be necessary to sum over the set of points that make up a single stimulus. Because of this, we will use the index $i$ to iterate over the points in a stimulus $D$ and omit $j$. The change with respect to the main text is given by:
\begin{align}
\textbf{x}^{(j)} \rightarrow \textbf{x} = \sum_{i=1}^N x_i \hat{e}_i \hspace{2cm} \textbf{y}^{(j)} \rightarrow \textbf{y} = \sum_{i=1}^N y_i \hat{e}_i.
\end{align}
During the derivations, we frequently use multiplications of the following form 
$\sum_{i=1}^N \boldsymbol{\phi}(x_i) \boldsymbol{\phi}^T(x_i) = \sum_{i=1}^N \boldsymbol{\phi}_i \boldsymbol{\phi}^T_i$,
where the result is a $D \times D$ matrix. To avoid confusion, we will always make the sum over the data points explicit with the running index $i$. We use bolt notation for vectors of dimensionality $D$, such as $\textbf{w}$ and $\boldsymbol{\phi}$  Finally, we use $\pi(\textbf{w})$ and $\rho(\textbf{w})$ as short-hands for the prior and posterior respectively. With these preliminaries, we are ready to work through the algebra for each of the three models.

\subsection{Maximum likelihood parameters}
ML-R uses on the likelihood maximizing parameter $\textbf{w}_{ML}$:
\begin{align}
p(y^\star | x^\star, D_j, \mathcal{M}_{\text{ML}}) &= p_{\sigma_g}(y^\star | x^\star, w_{ML}) \ \ 
 \text{with} \ \ w_{ML} = \arg \max_{w} p(D_j | w)
\label{eq:ML-prediction}
\end{align}
The maximizing parameter is:
\begin{align}
\label{eq:ML-explicit}
\textbf{w}_{ML} &= \left(\sum_{i=1}^N \boldsymbol{\phi}_i \boldsymbol{\phi}_i^T \right)^{-1} \sum_{i=1}^N y_i \boldsymbol{\phi}_i.
\end{align}
This follows from \cref{eq:ML-prediction} because we can find the maximum by setting the log-likelihood gradient to zero and solving for \textbf{w}:
\begin{align}
0 &\overset{!}{=} \partial_\textbf{w} \log \prod_{i=1}^N \mathcal{N} (y_i ; \textbf{w}^T\boldsymbol{\phi}(x_i),\sigma_g^{2}) \hspace{1cm} \Rightarrow \hspace{1cm} 0 = \sum_{i=1}^N (y_i - \textbf{w}^T\boldsymbol{\phi}_i) \boldsymbol{\phi}_i.
\end{align}
Solving this for $\textbf{w}$ yields \cref{eq:ML-explicit}.

\subsection{Maximum a posteriori parameters}
MAP-R uses the parameter that maximizes the posterior $p(w|D_j)= p(D_j | w)\pi(w)/ p(D_j)$:
\begin{align}
p(y^\star | & x^\star, D_j, \mathcal{M}_{\text{MAP}}) = p_{\sigma_g}(y^\star | x^\star, w_{MAP}) \ \ 
 \text{with} \ \ w_{MAP} = \arg \max_{w} p_{\sigma_g}(w|D_j).
\label{eq:MAP-prediction}
\end{align}
In order to find the global maximum of the posterior, we explicitly compute the posterior over the parameters $\textbf{w}$. This can be done in closed form because our mixed Gaussian prior is the conjugate prior of the Gaussian likelihood. Since the posterior is a Gaussian mixture again, there are only two potential candidates for the maximizing parameter, the means of the two components (indicated by the $\pm$ symbol):
\begin{align}
\label{eq:mu_rho-posterior}
\boldsymbol{\mu}_\rho^{(\pm)} &=  \Sigma^{(\pm)}_\rho \left( \sigma_g^{-2} \sum_{i=1}^N y_i \boldsymbol{\phi}_i +  (\Sigma_\pi^{(\pm)})^{-1} \boldsymbol{\mu}_\pi^{(\pm)} \right), \\
\label{eq:sig_rho-posterior}
\Sigma_\rho^{(\pm)} &:= \left((\Sigma_\pi^{(\pm)})^{-1} + \sigma_g^{-2} \sum_{i=1}^N \boldsymbol{\phi}_i \boldsymbol{\phi}_i^T \right)^{-1}.
\end{align}
To determine which of the two posterior modes is the global maximum, we numerically evaluate:
\begin{align}
\textbf{w}_{MAP} = \arg \max_{\xi = \pm} p(\boldsymbol{\mu}_\rho^{(\xi)}| D, \gamma).
\end{align}
To derive \cref{eq:mu_rho-posterior}, we start from the mixed prior:
\begin{align}
\pi(\textbf{w} | \gamma) = c_\pi^{(+)} \mathcal{N}\left(\textbf{w};\boldsymbol{\mu}_\pi^{(+)},\Sigma^{(+)}_\pi \right) + c_\pi^{(-)} \mathcal{N} \left(\textbf{w};\boldsymbol{\mu}_\pi^{(-)},\Sigma^{(-)}_\pi \right) = \sum_{\pm} c_\pi^{(\pm)} \mathcal{N} \left(\textbf{w};\boldsymbol{\mu}_\pi^{(\pm)} \Sigma^{(\pm)}_\pi \right),
\end{align}
where $\pm$ again denotes the two modes. After the presentation of a stimulus $D$ consisting of $N$ points, we show that the posterior, indexed by $\rho$ rather than $\pi$, resumes the form of a mixed Gaussian again:
\begin{align}
\rho(\textbf{w} |D, \gamma) &= \frac{p(D | \textbf{w}) \pi(\textbf{w} | \gamma)}{p(D|\gamma)}  =: \frac{1}{p(D|\gamma)} \sum_{\pm} c_\pi^{(\pm)} C^{(\pm)}  \notag \\
&=\frac{1}{p(D|\gamma)} \sum_{\pm} c_\pi^{(\pm)} \prod_{i=1}^N\mathcal{N}(y_i; \textbf{w}^T\boldsymbol{\phi}_i,\sigma_g^2) \mathcal{N}\left(\textbf{w};\boldsymbol{\mu}_\pi^{(\pm)},\Sigma^{(\pm)}_\pi \right) \notag \\
&=:
\sum_{\pm} c_\rho^{(\pm)} \mathcal{N}\left(\textbf{w};\boldsymbol{\mu}_\rho^{(\pm)}.\Sigma^{(\pm)}_\rho \right). \label{eq:posterior-update-abstract}
\end{align}
The symbol $C^{(\pm)}$ represents the terms in each component that contain the gist of the following algebraic manipulations. To get explicit expressions for the parameters of the posterior, i.e., $c_\rho^{(\pm)},\boldsymbol{\mu}_\rho^{(\pm)},\Sigma^{(\pm)}_\rho$, we evaluate the product of two Gaussians in each summand individually. To this end, we use completion of the square in the exponent $\chi$, introduced shortly. For a concise notation we temporally omit the index $\pm$. Copying from the second line in \cref{eq:posterior-update-abstract}, the important terms in each component read:
\begin{align}
C:= \left(\prod_{i=1}^N \mathcal{N}(y_i; \textbf{w}^T\boldsymbol{\phi}_i,\sigma_g^2)\right) \mathcal{N}\left(\textbf{w};\boldmath{\mu}_\pi,\Sigma_\pi \right) = (2 \pi \sigma_g^2)^{-N/2} |2 \pi \Sigma_\pi|^{-1/2} \exp \left(-\tfrac{1}{2} \chi \right). \label{eq:individual-product-of-two-gaussians}
\end{align}
To complete the square, we first group terms in powers of \textbf{w}:
\begin{align}
\chi &= \textbf{w}^T \left( \Sigma_\pi^{-1} + \sigma^{-2}_g \sum_{i=1}^N \boldsymbol{\phi}_i \boldsymbol{\phi}_i^T \right) \textbf{w} - \textbf{w}^T \left( \Sigma_\pi^{-1} \boldsymbol{\mu}_\pi + \sigma_g^{-2} \sum_{i=1}^{N} \boldsymbol{\phi}_i y_i \right) \notag  \\ \notag
& - \left( \Sigma_\pi^{-1} \boldsymbol{\mu}_\pi + \sigma_g^{-2} \sum_{i=1}^{N} \boldsymbol{\phi}_i y_i \right)^T \textbf{w} + \sigma_g^{-2} \sum_{i=1}^{N} y_i^2 + \boldsymbol{\mu}_{\pi}^T \Sigma^{-1}_\pi \boldsymbol{\mu}_{\pi} \\
&= (\textbf{w} - \boldsymbol{\mu}_\rho)^T \Sigma_\rho^{-1} (\textbf{w} - \boldsymbol{\mu}_\rho)  - \boldsymbol{\mu}_\rho^T \Sigma_\rho^{-1} \boldsymbol{\mu}_\rho + \sigma_g^{-2} \sum_{i=1}^{N} y_i^2 + \boldsymbol{\mu}_{\pi}^T \Sigma^{-1}_\pi \boldsymbol{\mu}_{\pi},
\label{eq:posterior-exponent}
\end{align}
where we already reported the result in terms of the posterior mean and variance (\cref{eq:mu_rho-posterior,eq:sig_rho-posterior}). Note that only the first term in the last line of \cref{eq:posterior-exponent} depends on $\textbf{w}$. It represents the exponent of one of the (Gaussian) mixture components of the posterior. The remaining terms go into the update of the mixture coefficients $c^{(\pm)}_\pi \rightarrow c^{(\pm)}_\rho$.
Substituting the simplified exponent \cref{eq:posterior-exponent} back into \Cref{eq:individual-product-of-two-gaussians} and adding a factor of $|2\pi \Sigma_\rho|$ to account for the normalization of the Gaussian mixture component, we have:
\begin{align}
C = \left( \frac{|\Sigma_\rho|}{(2 \pi \sigma_g^2)^N |\Sigma_\pi|} \right)^{1/2} \exp \left(-\tfrac{1}{2} ( \boldsymbol{\mu}_\rho^T \Sigma_\rho^{-1} \boldsymbol{\mu}_\rho - \sigma_g^{-2} \sum_{i=1}^{N} y_i^2 - \boldsymbol{\mu}_{\pi}^T \Sigma^{-1}_\pi \boldsymbol{\mu}_{\pi}) \right) \mathcal{N}(\textbf{w};\boldsymbol{\mu}_\rho, \Sigma_\rho).
\end{align}
Comparing this expression with \Cref{eq:posterior-update-abstract}, we identify the mixing coefficients of the posterior as $C$ without the $\textbf{w}$ dependent factor:
\begin{align}
\label{eq:c_rho-update}
\tilde{c}_\rho = c_\pi \left( \frac{|\Sigma_\rho|}{(2 \pi \sigma_g^2)^N |\Sigma_\pi|} \right)^{1/2} \exp \left(-\tfrac{1}{2} ( \boldsymbol{\mu}_\rho^T \Sigma_\rho^{-1} \boldsymbol{\mu}_\rho - \sigma_g^{-2} \sum_{i=1}^{N} y_i^2 - \boldsymbol{\mu}_{\pi}^T \Sigma^{-1}_\pi \boldsymbol{\mu}_{\pi}) \right).
\end{align}
The normalization then follows from the evidence term in \Cref{eq:posterior-update-abstract}. Now, labeling the two components of the posterior explicitly with $\pm$ again, we compute the evidence:
\begin{align}
p(D|\gamma) = \int \sum_{\pm} c_\pi^{(\pm)} C^{(\pm)} d\textbf{w} =
\sum_{\pm} \tilde{c}_\rho^{(\pm)} \int \mathcal{N}\left(\textbf{w};\boldsymbol{\mu}_\rho^{(\pm)},\Sigma^{(\pm)}_\rho \right) d\textbf{w} = \sum_{\pm} \tilde{c}_\rho^{(\pm)},
\label{eq:c_rho-posterior}
\end{align}
such that $c_\rho^{(\pm)} = \tilde{c}_\rho^{(\pm)} \left(\tilde{c}_\rho^{(+)} + \tilde{c}_\rho^{(-)} \right)^{-1} $ as expected.\\
With \cref{eq:mu_rho-posterior,eq:sig_rho-posterior,eq:c_rho-posterior} we have explicit expressions for the parameters of the mixed Gaussian as a result from combining the data points $D$ with the prior. This enables us to efficiently update the prior. Because the mixed Gaussian is a conjugate prior for the Gaussian likelihood, we the posterior maintains the form of a mixed Gaussian:
\begin{align}
\Sigma_\pi^{(\pm)} \rightarrow \Sigma_\rho^{(\pm)},
\hspace{2cm}
\boldsymbol{\mu}_\pi^{(\pm)} \rightarrow \boldsymbol{\mu}_\rho^{(\pm)},
\hspace{2cm}
c_\pi^{(\pm)} \rightarrow c_\rho^{(\pm)} = \frac{\tilde{c}_\rho^{(\pm)}}{\sum_{\pm} \tilde{c}_\rho^{(\pm)}}.
\end{align}

\subsection{Posterior predictive distribution}
B-R uses the entire posterior for making predictions by marginalizing over it:
\begin{align}
p(y^\star | & x^\star, D_j, \mathcal{M}_{\text{BR}})  = \int  p_{\sigma_g}(y^\star | x^\star, w) p_{\sigma_g}(w|D_j) dw \ \ \text{with} \ \ p_{\sigma_g}(w|D_j) = p_{\sigma_g}(D_j | w)\pi(w) p^{-1}(D_j)
\label{eq:BR-prediction}
\end{align}
To get an explicit expression for the posterior predictive in Bayesian regression, we have to solve the integral in \cref{eq:BR-prediction}: a posterior weighted average of forward predictions. The posterior predictive will be evaluated at the location $x^\star$ and we use $\boldsymbol{\phi}^\star:=\boldsymbol{\phi}(x^\star)$. To reduce notation, the poster predictive variable is simply denoted as $y$ (instead of $y^\star$ in the main text).\\
The algebraic steps to go from posterior to poster predictive, are similar to those of the previous section where we moved from prior to posterior. Again, the resulting expression is a mixed Gaussian and we adopt a similar notation as in the previous section, using the subindex $\beta$ (instead of $\rho$ for the posterior or $\pi$ for the prior) to indicate the posterior predictive. As before with $C$, we introduce the symbol $B$ as a shorthand for the relevant terms:
\begin{align}
p(y|x^\star,D,\mathcal{M}_{BR})
&=  \sum_{\pm} c_\rho^{(\pm)} \int \mathcal{N}(y ; \textbf{w}^T\boldsymbol{\phi}^\star, \sigma_g^2) \mathcal{N}(\textbf{w};\boldsymbol{\mu}^{(\pm)}_\rho, \Sigma^{(\pm)}_\rho) d \textbf{w} \\
&=: \sum_{\pm} c_\rho^{(\pm)} B^{(\pm)} =:  \sum_{\pm} c_\beta^{(\pm)}
\mathcal{N}\left(y;\mu_\beta^{(\pm)},\Sigma^{(\pm)}_\beta \right),
\label{eq:ppd-abstract}
\end{align}
Note that $\mu_\beta$ is a one-dimensional quantity, therefore not written as bold symbol. As before, we explicitly evaluate one of the two Gaussian mixture components in $y$ by rearranging the exponent. Omitting the label $\pm$ and indicating all auxiliary variable by a wiggle, we copy one of the two mixture components in \cref{eq:ppd-abstract}:
\begin{align}
B = c_\rho \int \mathcal{N}(y; \textbf{w}^T\boldsymbol{\phi}^\star,\sigma_g^2) \mathcal{N}\left(\textbf{w};\mu_\rho,\Sigma_\rho \right) d \textbf{w} =
c_\rho (2 \pi \sigma_g^2)^{-1/2} (|2 \pi \Sigma_\rho|)^{-1/2} \int \exp \left(-\tfrac{1}{2} \chi_0 \right) d \textbf{w},
\label{eq:individual-product-of-two-gaussians-in-ppd}
\end{align}
where the exponent $\chi_0$ contains terms in $\textbf{w}$ and $y$. We first group terms in powers of $\textbf{w}$ and rearrange them by completion of the square to result in a Gaussian that we integrate out (only the normalization constant remains):
\begin{align}
\chi_0 &= - \textbf{w}^T \left( \Sigma_\rho^{-1} + \sigma^{-2}_g \boldsymbol{\phi}_\star \boldsymbol{\phi}_\star^T \right) \textbf{w} - \textbf{w}^T \left( \Sigma_\rho^{-1} \boldsymbol{\mu}_\rho + \sigma_g^{-2}  \boldsymbol{\phi}_\star y \right) \\ \notag
& - \left( \Sigma^{-1} \boldsymbol{\mu}_\rho + \sigma_g^{-2} \boldsymbol{\phi}_\star y \right)^T \textbf{w} + a_g y^2 + \boldsymbol{\mu}_{\rho}^T \Sigma^{-1}_\rho \boldsymbol{\mu}_{\rho} \\ \notag
&= (\textbf{w} - \tilde{\boldsymbol{\mu}})^T \tilde{\Sigma}^{-1} (\textbf{w} - \tilde{\boldsymbol{\mu}})  - \tilde{\boldsymbol{\mu}}^T \tilde{\Sigma}^{-1} \tilde{\boldsymbol{\mu}} + a_g y^2 + \boldsymbol{\mu}_{\rho}^T \Sigma^{-1}_\rho \boldsymbol{\mu}_{\rho}.
\end{align}
This step was essentially similar to updating a Gaussian mixture with the data $d^\star = (x^\star,y)$. The new mean and variance appear as auxiliary parameters (marked with a wiggle):
\begin{align}
\tilde{\Sigma} &= \left( \Sigma_\rho^{-1} + \sigma^{-2}_g \boldsymbol{\phi}_\star \boldsymbol{\phi}_\star^T \right)^{-1}\\
\tilde{\boldsymbol{\mu}} &= \tilde{\Sigma} \left( \Sigma_\rho^{-1} \boldsymbol{\mu}_\rho + a_g \boldsymbol{\phi}_\star^T y \right).
\end{align}

Now we return to \Cref{eq:individual-product-of-two-gaussians-in-ppd} and carry out the marginalization, which amounts to a factor of $|2\pi\tilde{\Sigma}|^{1/2}$:
\begin{align}
B = 
 \left( \frac{|\tilde{\Sigma}|}{2 \pi \sigma_g^2 |\Sigma_\rho|}\right)^{1/2} \exp \left(-\tfrac{1}{2}\chi_1 \right). \label{eq:individual-product-of-two-gaussians-in-ppd2}
\end{align}
Next we turn to the reduced exponent $\chi_1$, group terms in powers of $y$ and complete the square to identify the mixture component of the posterior predictive (a Gaussian in $y$):
\begin{align}
\chi_1 &= - \tilde{\boldsymbol{\mu}}^T \tilde{\Sigma}^{-1} \tilde{\boldsymbol{\mu}} + \sigma_g^{-2} y^2 + \boldsymbol{\mu}_{\rho}^T \Sigma^{-1}_\rho \boldsymbol{\mu}_{\rho} \notag \\
&=  (\sigma_g^{-2} - \sigma_g^{-2} \boldsymbol{\phi}_{\star}^T \tilde{\Sigma} \boldsymbol{\phi}_{\star} \sigma_g^{-2} )y^2 + 2 (\sigma_g^{-2} \boldsymbol{\phi}_\star^T \tilde{\Sigma} \Sigma_\rho^{-1} \boldsymbol{\mu}_\rho ) y - \boldsymbol{\mu}_\rho^T \Sigma^{-1}_\rho \tilde{\Sigma} \Sigma^{-1}_\rho \boldsymbol{\mu}_\rho + \boldsymbol{\mu}_{\rho}^T \Sigma^{-1}_\rho \boldsymbol{\mu}_{\rho} \notag \\
&=: \sigma^{-2}_\beta  y^2 + 2 \sigma^{-2}_\beta \mu_\beta   - \boldsymbol{\mu}_\rho^T \Sigma^{-1}_\rho \tilde{\Sigma} \Sigma^{-1}_\rho \boldsymbol{\mu}_\rho  + \boldsymbol{\mu}_{\rho}^T \Sigma^{-1}_\rho \boldsymbol{\mu}_{\rho} \notag \\
&=: \sigma^{-2}_\beta(y - \mu_\beta)^2 + \chi_2,
\end{align}
where in the second last step we identified the mean and precision of the posterior predictive and in the last step we gave the reduced constant a new name after completion of the square:
\begin{align}
\sigma^{2}_\beta &= \left( \sigma^{-2}_g - \sigma^{-2}_g \boldsymbol{\phi}_{\star}^T \tilde{\Sigma} \boldsymbol{\phi}_{\star} \sigma^{-2}_g \right)^{-1} \label{eq:ppd-precision-update} \\
\mu_\beta &= \frac{\sigma^{2}_\beta}{\sigma_g^{2}} \boldsymbol{\phi}_{\star}^T \tilde{\Sigma} \Sigma_\rho \boldsymbol{\mu}_\rho \label{eq:ppd-mean-update} \\
\chi_2 &:= - a_\beta \mu_\beta^2 - \boldsymbol{\mu}_\rho^T \Sigma^{-1}_\rho \tilde{\Sigma} \Sigma^{-1}_\rho \boldsymbol{\mu}_\rho + \boldsymbol{\mu}_{\rho}^T \Sigma^{-1}_\rho \boldsymbol{\mu}_{\rho}
\end{align}
Substituting this into \Cref{eq:individual-product-of-two-gaussians-in-ppd} and adding the normalization constant of the Gaussian over in $y$ yields for s single mixture component:
\begin{align}
B = 
 \left(  \frac{\sigma^2_\beta|\tilde{\Sigma}|}{\sigma^2_g |\Sigma_\rho|}\right)^{1/2}  \exp(-\tfrac{1}{2}\chi_2 ) \mathcal{N}(y; \mu_\beta, a_\beta^{-1})
\end{align}
The update of the mixing constants follows from the pre-factors and the reduced exponent:
\begin{align}
c_\beta \propto c_\rho \left(\frac{\sigma^2_\beta | \tilde{\Sigma} | }{\sigma^2_g |\Sigma_\rho|}  \right)^{1/2}  \exp(-\tfrac{1}{2}\chi_2 ), \label{eq:ppd-mixing-update}
\end{align}
where the proportionality indicates that the coefficients on the right hand side must still be normalized to one. Together \Cref{eq:ppd-precision-update,eq:ppd-mean-update,eq:ppd-mixing-update} report how to compute posterior predictive in \Cref{eq:ppd-abstract}.

\subsection{Prior predictive}
As a null model, we tested if the measured responses could be explained simply in terms of the prior predictive: 
\begin{align}
\label{eq:Prior-prediction}
p(y^\star | x^\star, D_j, \mathcal{M}_{PR})  &= \int p(y^\star | x^\star, w) \pi(w) dw
\end{align}
Computing the prior predictive requires exactly the same steps as in the previous section. The only difference is that our starting point is not posterior but the prior. In terms of the equations, this amounts to replacing all subindices $\rho$ (posterior) with the subindex $\pi$ (prior) in the previous section.

\subsection{Derivation of full Bayesian regression}
\label{sec:supp:BRf}
To compute the posterior predictive distribution, we can expand on our previous results for ordinary B-R, namely the fact that know how to compute 
\begin{align}
p(y^\star | x^\star, D_j, \mathcal{M}_{\text{BR}})  = \int p(y^\star | x^\star, w, \sigma_g')  p(w|D_j) \text{d}w 
\label{eq:supp:BR->BRf}
\end{align}
for any fixed non-zero value of $\sigma_g'$. To extend to full B-R, we make two modifications in \cref{eq:supp:BR->BRf}. First, we replace the posterior by the joint posterior:
\begin{align}
    p_{\sigma_g'}(w |D_j) = p_{\sigma_g'}(D_j | w) \pi(w) p^{-1}(D_j)  \rightarrow p_{\sigma_g'}(D_j | w) \pi(w) \pi(\sigma_g') p^{-1}(D_j) = p(w, \sigma_g'|D_j).
\end{align}
Note that the normalization constant is not the same on the LHS and RHS of the equation. The only change is the factor of $p_{\sigma_g'}$ and the fact that we have no closed form expression for the normalization anymore. Second, we integrate over $\sigma_g'$ in \cref{eq:supp:BR->BRf}. Because we lost the closed form expression for the normalization constant, we use proportionality instead of equality:
\begin{align}
p(y^\star | x^\star, D_j, \mathcal{M}_{\text{BR}_f}) &= \int p(y^\star | x^\star, w, \sigma_g')  p(w, \sigma_g'|D_j) \text{d}w \text{d}\sigma_g' \\ 
&=
 \int p(y^\star | x^\star, w, \sigma_g')  p_{\sigma_g'}(D_j | w) \pi(w) \pi(\sigma_g') p^{-1}(D_j) \text{d}w \text{d}\sigma_g' \\ 
& \propto  \int \pi(\sigma_g') p(y^\star | x^\star, D_j, \mathcal{M}_{\text{BR}})  \text{d}\sigma_g'
\label{eq:supp:BRf-derivation}
\end{align}
where we dropped the normalization constant because it is numerically more convenient to normalize only the final expression, ie the posterior predictive distribution. Note that we owe the simplicity of this extension to the fact that chose a factorized prior over $\pi(w,\sigma'_g)=\pi(w)\pi(\sigma'_g)$. When considering sequential inference, this choice has the clear downside that the prior is not conjugate to the gaussian likelihood. However, this does not come to bear in our work because we consider only one inference step upon seeing the stimulus.\\
\\
We choose $\pi(\sigma'_g)$ as an exponential distribution. We use the data generating hyperparameter $\sigma_g$ to fix the expectation of the  such that $\pi(\sigma'_g)$:
\begin{align}
    \text{E}[\pi(\sigma'_g)] \overset{!}{=} \sigma_g  \ \ \Rightarrow \ \ \pi(\sigma'_g) = \sigma_g^{-1} \exp(-\sigma'_g / \sigma_g).
\end{align}

\subsection{Bayesian regression with three and four-point estimates of the generative noise}
\label{sec:supp:BR3}
\begin{figure}[h]
\centering
\begin{minipage}{0.45\textwidth}
\includegraphics[width=\textwidth]{./BR+3.pdf}
\end{minipage}
\begin{minipage}{0.45\textwidth}
\includegraphics[width=\textwidth]{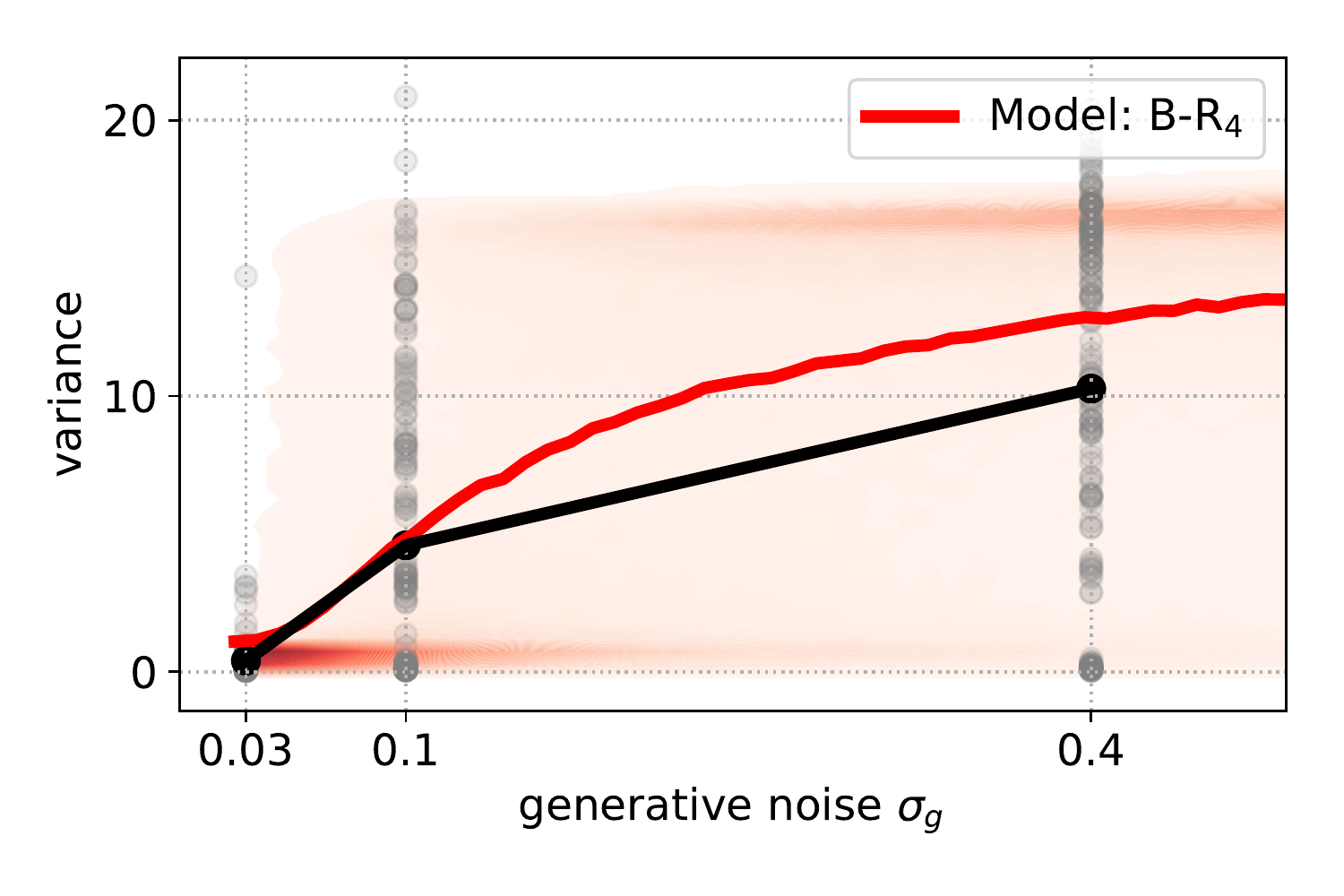}
\end{minipage}
\caption{ \label{fig:supp:br+4}
\textbf{(Left)} Variance distribution for B-R$_3$ and data, same plot as Fig. 4 in the main text (bottom, right). \textbf{(Right)} Variance distribution for B-R$_4$. Qualitatively it does not differ from B-R$_3$ but it does not capture the mean (black) variance in the data and the fraction of almost zero responses.
}
\end{figure}

To estimate the generative noise, from $M \in \{3,4\}$ points, we set the gradient of the log-likelihood to zero. The closed from solution for the estimate of the generative noise is:
\begin{align}
\hat{\sigma_g}^2 = M^{-1} \sum_{i=1}^M (y_i - w^\star x_i^2)^2 \ \ \text{with} \ \ w^\star = \left(\sum_{i=1}^M x_i^4 \right)^{-1} \sum_{i=1}^M y_i x_i^2.
\end{align}
For the B-R$_3$ method, we compute two such estimate by ignoring the left and right most stimulus point and then take the minimum value as input for Bayesian regression. For the B-R$_4$ method used to generate \cref{fig:supp:br+4} we used $M=4$ to obtain a single noise estimate.\\
\\
While both methods B-R$_3$ and B-R$_4$ are heuristics, we opted for B-R$_3$ because it better captures the empirical variance distribution, as confirmed by \cref{fig:supp:br+4}.

\subsection{Bayesian regression scales with generative noise}
\label{sec:noise-limits}
Here, we examine the limiting cases of $\sigma_g \rightarrow 0$ and $\sigma_g \rightarrow \infty$. We show that all three models (ML, MAP and BR) make the same prediction in the former case, and that the posterior predictive and the prior prediction collapse in the latter case. Both statements are shown by examining the limiting cases of the posterior. It is not necessary to work with the posterior predictive.\\
\subsubsection*{Small generative noise}
Based on the data $D = (x_j,y_j)_{j=1}^{N}$, ML makes a prediction with based on $\textbf{w}_{ML}$ (\cref{eq:ML-explicit}). Note that this prediction is independent of $\sigma_g$. The MAP prediction is given by \cref{eq:mu_rho-posterior,eq:sig_rho-posterior}:
\begin{align*}
\boldsymbol{\mu}_\rho^{(\pm)} &=  \left((\Sigma_\pi^{(\pm)})^{-1} + \sigma_g^{-2} \sum_{i=1}^N \boldsymbol{\phi}_i \boldsymbol{\phi}_i^T \right)^{-1} \left( \sigma_g^{-2} \sum_{i=1}^N y_i \boldsymbol{\phi}_i + (\Sigma_\pi^{(\pm)})^{-1} \boldsymbol{\mu}_\pi^{(\pm)} \right) \\
&= \left(\sigma_g^{2} (\Sigma_\pi^{(\pm)})^{-1} + \sum_{i=1}^N \boldsymbol{\phi}_i \boldsymbol{\phi}_i^T \right)^{-1} \left( \sum_{i=1}^N y_i \boldsymbol{\phi}_i + \sigma_g^{2} (\Sigma_\pi^{(\pm)} )^{-1} \boldsymbol{\mu}_\pi^{(\pm)} \right) \overset{\sigma_g \rightarrow 0}{\rightarrow} \textbf{w}_{ML},
\end{align*}
as the data term dominates the prior. This makes intuitive sense, if the data is highly informative we do not rely on the prior. \\
The same holds true for Bayesian regression because the bimodal posterior becomes a delta function around the ML-solution. We already established that the mean associated with each of the two modes converge to the maximum likelihood solution, i.e.,  $\boldsymbol{\mu}_\rho^{(\pm)} \rightarrow \textbf{w}_{ML}$. We also know that the mixture coefficients $c^{(\pm)}_\rho$ always sum to one. Now, we only have to verify that the covariance matrices of both modes converge to zero:
\begin{align}
\Sigma_\rho^{(\pm)} = \sigma^2_g \left( \sigma^2_g (\Sigma_\pi^{(\pm)})^{-1} + \sum_{i=1}^N \boldsymbol{\phi}_i \boldsymbol{\phi}_i^T \right)^{-1} \overset{\sigma_g \rightarrow 0}{\rightarrow} "0 \left( \sum_{i=1}^N \boldsymbol{\phi}_i \boldsymbol{\phi}_i^T \right)^{-1}" = 0.
\end{align}
With a delta-peaked posterior, the BR prediction collapses into the ML prediction as well:
\begin{align}
p(y|x,D,\mathcal{M}_{BR}) = \int \delta(\textbf{w} - \textbf{w}_{ML}) p(y|x,\textbf{w}) \text{d}\textbf{w} = p(y|x,\textbf{w}_{ML}).
\end{align}
Therefore all three methods make the same prediction.

\subsubsection*{Large noise limit}
In the limit of $\sigma_g \rightarrow \infty$, the data carries no information and the posterior (indicated by $\rho$) distribution converges to the prior distribution (indicated by $\pi$). Then BR makes the same prediction as a prediction based on the prior. It is straightforward to show that the mean and variance of each mode converge to the prior's mean and variance:
\begin{align}
\label{eq:large-noise-limit-sig}
\Sigma_\rho^{(\pm)} &= \left(  (\Sigma_\pi^{(\pm)} )^{-1} + \sigma^{-2}_g \sum_{i=1}^N \boldsymbol{\phi}_i \boldsymbol{\phi}_i^T \right)^{-1} \overset{\sigma_g \rightarrow \infty}{\rightarrow} \Sigma_\pi^{(\pm)} \\
\label{eq:large-noise-limit-mu}
\boldsymbol{\mu}_\rho^{(\pm)} &= \Sigma_\rho^{(\pm)} \left( \sigma_g^{-2} \sum_{i=1}^N y_i \boldsymbol{\phi}_i +  (\Sigma_\pi^{(\pm)})^{-1} \boldsymbol{\mu}_\pi^{(\pm)} \right) \overset{\sigma_g \rightarrow \infty}{\rightarrow} = \boldsymbol{\mu}_\pi^{(\pm)}.
\end{align}
As a final step, we have to verify that the coefficients of the modes do not change. \Cref{eq:c_rho-update} prescribes the update of the coefficients. The update can be expressed by the update factor $f^{(\pm)}$ between the normalized prior coefficient $c_\pi^{(\pm)}$ and the unnormalized posterior coefficient $\tilde{c}_\rho^{(\pm)}$ is: $f^{(\pm)} c_\pi^{(\pm)} = \tilde{c}_\rho^{(\pm)}$. If the update factor is independent the original mode, i.e., independent of the label $\pm$, both mixture components receive the same update and since the final coefficients are normalized, this amounts to no update at all. It is easy to see that in the limit of $\sigma_g \rightarrow \infty$, the update becomes independent of original mode indeed:
\begin{align}
\label{eq:c_rho-update}
f^{(\pm)} = \left( \frac{|\Sigma_\rho^{(\pm)}|}{(2 \pi \sigma_g^2)^N |\Sigma_\pi^{(\pm)}|} \right)^{1/2} \exp \left(-\tfrac{1}{2} ( [\boldsymbol{\mu}_\rho^T \Sigma_\rho^{-1} \boldsymbol{\mu}_\rho]^{(\pm)} - \sigma_g^{-2} \sum_{i=1}^{N} y_i^2 - [ \boldsymbol{\mu}_{\pi}^T \Sigma^{-1}_\pi \boldsymbol{\mu}_{\pi}]^{(\pm)} )  \right)
\end{align}
With \cref{eq:large-noise-limit-sig,eq:large-noise-limit-mu} it follows directly that $f^{(\pm)}$ becomes the same for both modes. Therefore, prior and posterior are equivalent.

\FloatBarrier




\dataset{subjN\_sig\_g=X.txt}{The name of the data files contains to placeholders $N$ and $X$. They enumerate the participants $N \in \{1, \dots 7\}$ and indicate the generative noise level $X \in \{0, 0.03, 0.1,0.4\}$ respectively. For example, the file \texttt{subj1\_sig\_g=0.1.txt} contains all trials of the first participant with generative noise $\sigma_g=0.1$. Each data file contains 11 columns. The first eight columns describe the $x$ and $y$ coordinates of the stimulus points. The last three columns contain (in that order) the stimulus index $j \in \{1, \dots 20 \}$, the generating quadratic parameter $w_j$ and the vertical location of the measured response.}

